\documentclass[12pt]{article}
\usepackage{epsf}
\usepackage{cite}

\def\ber{\begin{eqnarray}}
\def\eer{ \end{eqnarray}}  
\def\be{\begin{equation}}
\def\ee{\end{equation}}
\def\la{\label} 
\def\lan{\langle} 
\def\ran{\rangle} 
\def\lb{\lbrace} 
\def\rb{\rbrace} 
\def\lbr{\lbrack} 
\def\rbr{\rbrack} 
\def\ep{E_+}
 \def\zp{{z_+ }}
 \def\cp{{\overline{z}_+ }}
 \def\zm{{z_- }}
 \def\cm{{\overline{z}_- }}
 \def\cz{\overline{z} }
 \def\CZ{\overline{Z}}
 \def\ZP{Z_+ }
 \def\CP{\overline{Z}_+ }
 \def\ZM{Z_- }
 \def\CM{\overline{Z}_- }
 \def\zpm{{z_{\pm} }}
 \def\ZPM{{Z_{\pm} }}
 \def\CPM{{\overline{Z}_{\pm} }}
 \def\cpm{{\overline{z}_{\pm} }}

 \def\hzp{{\hat{z}_+ }}
 \def\hcp{{\hat{\overline{z}}_+ }}
  \def\hzm{{\hat{z}_- }}
 \def\hcm{{\hat{\overline{z}}_- }}
 \def\pzp{{\frac{\partial}{\partial{z_+} }}}
 \def\pcp{{\frac{\partial}{\partial{\overline{z}_+} }}}

 \def\hzpm{{\hat{z}_\pm}}
 \def\hcpm{{\hat{\overline{z}}_\pm }}
 \def\hf {{\frac{1}{2}}}
 
 \def\n{\nonumber\\}

 \def\b{\beta}
 
 \def\nR{ {\frac{n}{R}} }

 \def\f{\frac}
\font\litfont = cmr6

\def\bigone{\hbox{1\kern -.23em {\rm l}}}     
\def\ZZ{\hbox{\zfont Z\kern-.4emZ}}
\def\hf{{\litfont {1 \over 2}}}

\def\p{\partial}

\def\b{\beta}

\def\d{\delta}

\def\m{\mu}
\def\al{\frac{1}{\sqrt{2\alpha^\prime}}}
\def\cz{\overline{z}}

\def\vp{\varphi}

\def\O{\Omega}
\def\oo{\hat \omega   }

\def\o{\omega }

  \def\tpr{2{\pi}R} 
  
  \def\CF {{\cal F}}

  \def\CN {{\cal N}}

  \def\CR {{\cal R}}
  
  \def\CT {{\cal T}}

  \def\CW {{\cal W}}

\def\Pip#1{\Psi^{_{#1}}_{+}}
\def\Pim#1{\Psi^{_{#1}}_{-}}

\def\o{\omega}

\catcode`@=11 \@addtoreset{equation}{section} \catcode`@=12

\begin{document}

\setlength \arraycolsep{2pt}

\begin{titlepage}
\vfill
\begin{center}
{\Large \bf On Integrability of   Type 0A Matrix model in the presence  of D brane}\\[1cm] 
Chandrima Paul {\footnote{mail:plchandrima@gmail.com}} \\ 
\vskip3mm
\emph{ Department of Physics, \\Sikkim University, 6th Mile, Gangtok },\\
\emph{Sikkim 737102, India\\[10pt]}
\end{center}
\vfill

\begin{abstract} 
We consider type 0A matrix model in the presence of spacelike D brane which is localized in matter direction at any arbitrary point.  In string theory, the boundary state which in matrix model corresponds to the Laplace transform of the macroscopic loop operator, is known to obey the operator constraints corresponding to open string boundary condition. When we analyze   MQM as well as the respective collective field theory and compare it with dual string theory it appears that consistency of the theory requires a condition  equivalent to a constraint on the matter part that needed to be imposed in the matrix model.  We identified this condition and observed that this has only effect into constraining the macroscopic loop operator so that it projects the Hilbert space generated by the operator to its physical sector at the point of insertion while keeping  the bulk matrix model remains unaffected, thereby describing a situation parallel to string theory. We analyzed the theory with uncompactified time and have shown explicitly that the matrix model predictions are in  good agreement with the relevant string theory. Next we considered the theory with compactified time, analyzed  MQM on a circle in the presence of  D brane.  We evaluated the partition function along with the constrained macroscopic loop operator in the grand canonical ensemble and  showed the free energy corresponds to that of a deformed Fermi surface.   We have also shown that the path integral in the presence of D brane can be expressed as the Fredholm determinant. We have studied the fermionic scattering in a semiclassical regime.  Finally we considered the compactified theory in the presence of the D brane with tachyonic background.  We evaluated the free energy in the grand canonical ensemble. We have shown the integrable structure of the respective partition function and it corresponds to the tau function of Toda hierarchy.  We have also analyzed the dispersionless limit.{\footnote{Work is done in IIT Bombay}}
\end{abstract} 
\vfill 
\end{titlepage}

\begin{center}
\tableofcontents
\end{center}
\newpage

\section{Introduction}
\setcounter{equation}{0}

\subsection{Introduction to the matrix model in the presence of D brane : A review}

The two dimensional string theory (see e.g. \cite{KIR}, \cite{GM},
\cite{poreview} for reviews) is a very instructive model when we would
like to understand the nature of string theory as a complete
theory of quantum gravity. This theory has a powerful dual
description of $c=1$ matrix model defined by the simple quantum
mechanics of a Hermitian matrix $\Phi$ with the inverse harmonic oscillator
potential $U(\Phi)=-\Phi^2$ after the double scaling limit.Matrix model is successfully used to
 describe 2D string theory in the simplest linear dilaton background as well as to incorporate
perturbations.
\vskip0.5mm
 In last decade  the $c=1$ matrix quantum mechanics has received 
lots of attention because of its new interpretation as the
decoupled world volume theory of unstable
D0-branes \cite{Reloaded,clasquadbrn,KMSdecay}. The matrix model
dual to type 0 string theories were also proposed in
\cite{TadashiNick,newhat}.  In particular, the matrix model dual of the two dimensional type 0
string  gives a non-perturbatively well-defined
formulation. For example, the type
0B model is defined by the hermitian matrix model with two Fermi
surfaces. The type 0B matrix
quantum mechanics (MQM) describes open string tachyons living on
the unstable D0-branes, whereas the type 0A MQM describes
tachyonic open strings stretched between stable D0- and
anti-D0-branes. Upon compactification on Euclidean time, these two
matrix models are conjectured to be T-dual to each other. The
exact agreement in free energy was found in \cite{newhat}. Matrix model dual of type 0 string in the flux background was  explored in
 \cite{fluxvacua,fluxbgd}.  However, unlike $c=1$ matrix model which can be derived from discretizing
the Polyakov action on the string world sheet, such a derivation is
not known for type 0 matrix models.
\vskip.2mm
  An attempt  was made in \cite{taka} to obtain the exact form of the macroscopic loop operator in Type 0 string theory. If we consider the bosonic string  partition function
\be
\int D{\phi}DX \,\,exp \left\lbrack -\int {d^2}z \lbrack\,{\frac{1}{4\pi}}( {\partial{ X}\partial{X}}+{\partial{\phi}\partial{\phi}})+QR{\phi}+{\mu}e^{2b\phi}\rbrack
-\,\int_{\partial\Sigma}{d\xi}\,\,\lbrack {\frac{Qk\phi}{2\pi}}+{\mu_B}e^{b\phi}\,\rbrack\,\, \right\rbrack ,
\la{bosonic}
\ee

the macroscopic loop operator inserts an operator
\be  
{W(t,l)\sim \delta\left(\int_{\partial\Sigma}e^{\phi}-l\right)\cdot \delta(X^0-t)},
\la{macroboson}
\ee

within the path integral \cite{loopstostates,loopstofields}.
\be
\lan W(l) \ran = Z(l) = \int DX D\phi D\lbr\,{\rm ghost}\,\rbr\,\, \delta\left(\int_{\partial\Sigma}e^{\phi}-l\right)\cdot \delta(X^0-t)\, f(x,\phi)\,Z(\phi(\sigma),X(\sigma),\lbr{\rm ghost}\rbr),
\la{mbosonic}
\ee
where f is some wave function for matter ghost and Liouville.
The physical meaning of this operator in two
dimensional string theory is the presence of a `Euclidean D-brane'
localized in the time direction. To be more precise
after we take the Laplace transformation $\int d\phi e^{-\mu_B
e^{\phi}}$, we get a D-brane with the Neumann boundary condition
 in the Liouville direction 
and the Dirichlet one in the time direction 
\be
\int
{\f{dl}{l}}e^{-\mu_B l}\ W_{bos}(t,l) \simeq
|B_{(FZZT)}(\mu_B)\ran_{\phi}\otimes |D\ran_{X^0}.
\la{op}
\ee
when we impose the  condition that boundary Liouville term is zero.
\be
{\p_n}{\phi} + {\mu_B}{e^{b\phi}} = 0
\la{lboundary}
\ee
Where ${\p_n}{\phi}$ denotes the Liouville momentum normal to boundary {while along the boundary we have ${\p_t}{X^o} = 0$.  
Now consider 2D superstring action  obtained from  extending the bosonic fields to their superspace and expanding the 2D superspace action in terms of the component field,
\be
S=
{\frac{1}{2\pi}}\int {d^2}z\lbrack {\delta_{\mu\nu}}
( {\partial} {X^\mu}{\overline{\partial}}{X^\nu} +
{\psi^\mu}{{\overline\partial}{\psi^\nu}}+
{\overline{\psi^\mu}}{\partial}{\overline{\psi^\nu}}) +{\frac{Q}{4}}R{X^1}\rbrack
+2i\mu{b^2}\int {d^2}z({ \psi^1}{\overline{\psi^1}}+2\pi\mu{e^{\phi}})
:e^{\phi}:
\la{superbosonic}
\ee
We can also consider the macroscopic loop operator which is the superspace analogue  of $W_{bos}(t,l)$, inserts the boundary condition on the fermionic coordinate ${\overline{\psi}} ( \overline{z}) = {\eta} {\psi}(z)$ where ${\eta}=\pm 1 $ describes the RR and NS NS sector.  Laplace transform of this operator inside the string path integral  describes the boundary states  NS NS and RR sector.   However depending on helicity, in each sector we have two types of boundary states
${\epsilon =\pm}$ so that we have four types of macroscopic loop operator given by ${ W_{NS}^+}, { W_{NS}^-}, { W_{R}^+},{ W_{R}^-}$.
 The parameter ${\mu_B}$ corresponds to the boundary cosmological constant in the
boundary state.  Indeed we can show this relation \cite{fzzt,zz} by
computing one point function on the brane or equally annulus amplitude as shown in \cite{annularreport}.  For $c=1$ matrix model the expression of these operators were obtained and its equivalence to string theory is verified in \cite{loopstostates,loopstofields,annularreport}.  Author of \cite{taka} obtained the expressions of  macroscopic loop operator in Type 0B matrix model and also for ${NS}$ sector of Type 0A matrix model 
which was verified by calculating the one point function.

Now once we understand the duality between noncritical string theory in the linear dilaton background and Matrix model,  its natural to ask whether we can understand the string theory with nontrivial background which has an obvious realization in matrix model by adding perturbations which survive in the double scaling 
limit. There are two ways to change the background of string theory: either to consider strings
propagating in a non-trivial target space or to introduce the perturbations .
In the first case one arrives at a complicated sigma-model. Not many examples are known
when such a model turns out to be solvable. Besides, it is extremely difficult to construct
a matrix model realization of a general sigma-model since not much known about matrix
operators explicitly perturbing the metric of the target space. Thus, we lose the possibility
to use the powerful matrix model machinery to tackle our problems.
    On the other hand, following the second way, we find that the integrability of the theory
in the trivial background is preserved by the perturbations. Also when we study the theory in a nontrivial background in most of the cases the target space metric of such
backgrounds is curved and often it incorporates the black hole
singularities.  In the superstring theories, the supersymmetry allows
for some interesting nontrivial solutions which are stable and
exact. But the string theory on such backgrounds is usually an
extremely complicated sigma-model, very difficult even to formulate it
explicitly, not to mention studying quantitatively its dynamics.
The two-dimensional bosonic string theory as well as Type 0 theory are the rare cases of sigma-model
where such a dynamics is integrable, at least for some particular
backgrounds, including the dilatonic black hole background.  A
physically transparent way to study the perturbative (one loop) string
theory around such a background is provided by the CFT approach.  However once  we try to understand  higher loops or multipoint correlators, we
have to address ourselves to the matrix model approach to the 2D
string theory . 
\vskip0.2mm
 The 2D	string theory has been constructed as
the collective field theory \cite{stringfield},\cite{lectjevicki}, in which
the only excitation, the massless tachyon, was related to the
eigenvalue density of the matrix field.  Now consequence of the deformation in eigenvalue density corresponding to deformation in string background at classical limit was studied in \cite{classicallimit}.  $C=1$ string theory perturbed by tachyonic mode studied in \cite{kostovmain}.  Vortex perturbation and its equivalence to sine -Liouville theory was studied in \cite{boulatov},\cite{matblackholekazakov}.  Its shown that partition function is integrable and have Toda structure. Toda structure and Lax formalism in the context of matrix model described in \cite{takasaki}.  Many more works in this direction  was done in \cite{yin,timedependentbgd,holography,openclosedual,nonperturbative,nonperturbativedbranes,thermodynamics,vortex}.

\subsection{Introduction to a new operator constraint in the matrix model in the presence of D brane and the integrability of the theory}

Now it is  an interesting question to ask that can we study this nontrivial background in the context of dual matrix model  in the presence of D brane which are just the Laplace transform of the macroscopic loop operator, as we discussed. Open close duality predicts that partition function must have integrable structure. 
           We are going to explain that the theory in the presence of D brane is not integrable in general.  This can be clearly shown from the expression of the macroscopic loop operator(as we are going to discuss in section 2.1, 2.2).  The reason that we fail to find the integrability is that, we need to impose one extra condition in matrix model  which was so far not observed.  Here in the introduction we very briefly mention the origin of such constraint. 
\vskip1mm 
Note,  we express the presence of D brane  by the insertion of the operator ${e^{\int dt W(t)\d(t-{t_o})} }$ in matrix model path integral , where W(t) have different expression in different matrix model.
Now, our observation in section 2.2, is that, in the presence of D brane operator ${e^{\int dt W(t)\d(t-{t_o})} }$ in the matrix model path integral, MQM hamiltonian is not conserved in general , i.e $\int_{t_o - \epsilon}^{t_o +\epsilon} dt\, \p_t \lan\,{ H_o} (t)\,\ran \ne 0$, where $t_o$ is the time of insertion of D brane operator( see \ref{leak}). Indeed the nonconservation equation is given by the expression $\int_{t_o - \epsilon}^{t_o +\epsilon} dt\, \p_t \lan\,{ H_o} (t)\,\ran + \d \lan \,W({t_o}) \,\ran = 0$, as we are going to prove in section 2.2.
\vskip1mm
Now here our claim is, matrix model hamiltonian must be conserved even in the presence of D brane operator in matrix path integral, so that
we put an extra constraint 
\be
(\delta_t W(t))|_{t= t_o} = 0. 
\la{initialcondition}
\ee
which confirm the conservation of $H_o$ in (\ref{leak}).
\vskip1mm
We are going to discuss in details, the way this constraint arise in matrix model in section 2.2.  We also discuss there that how it constrain the D brane operator in matrix model(not the MQM eigenstates!). We also discuss string theoretical meaning of the constraint in section 2.4.
\vskip1mm
Here in the introduction, we briefly show from MQM/String duality, why we need to put such constraint, if (\ref{leak}) is true. 
\vskip1mm
Now MQM is in one to one correspondence with string theory.  String theory Polyakov path integral sum, is equal to the logarithm of matrix model partition function, even in the presence of D brane operator \cite{KIR}.
\vskip1mm
Let us consider the Polyakov path integral sum in its hamiltonian representation. In order to understand that first we discuss the situation in ordinary QFT.  Note, in the ordinary QFT, the path integral sum $\int \prod d\phi(x) e^{-\int_i^f d^4 x L}$ (where the boundary condition i and f is such that the initial and final states are $|i\ran,|f\ran$)  in hamiltonian representation is given by $\lim\limits_{T\rightarrow\infty} \lan f| e^{-\int^T_{-T} dt H}|i\ran  $. These two are equal even with the insertion of any operator $O(t_o)$,i.e $ \int\prod d\phi(x) e^{-\int d^4 x L} O(t_o) = \lim\limits_{T\rightarrow\infty} \lan f| e^{-\int^T_{t_o} dt H} O(t_o)e^{-\int^{t_o}_{-T} dt H} |i\ran$ with a possible normalization. 
\vskip1mm
Similar equality one can write for string theory Polyakov path integral $$\displaystyle\sum_{\rm topologies} D[g_{\mu \nu}] D[X] e^{-S} =\lan f;\tau =\infty| e^{-\int H_{\rm string} d\tau}|i,\tau=-\infty\ran $$.   Here $H_{\rm string}$  is the string hamiltonian, generator of worldsheet time $\tau$ translation symmetry and, $|i;\tau = \infty \ran$,$|f;\tau=-\infty\ran$,represents  string states at infinite past and infinte future in world sheet time $\tau$.  Now this must be true even in the presence of insertion of D brane operator
\be
\displaystyle\sum_{\rm topologies} D[g_{\mu \nu}] D[X] e^{-S}O = \lan f;\tau =\infty| e^{-\int H_{\rm string} d\tau} O|i,\tau=-\infty\ran,
\la{rn}
\ee
 where O is the Laplace transform of macroscopic loop operator given in (\ref{op}). 
However, according to string/ MQM duality,  the L.H.S  of (\ref{rn})  is equal to the logarithm of matrix model partition function Z in the presence of D brane operator $e^{W(t_o)}$, where the matrix model partition function Z, in the presence of D brane operator, in hamiltonian representation, is given by   $Z=\lan f_M|  e^{\int_{t_o}^\infty dt H}e^{W(t_o)}e^{\int_{-\infty}^{t_o} dt H }|i_M\ran$  where $|i_M\ran$ and $|f_M\ran$ are matrix model states at infinite past and future.
\vskip1mm
Finally the meaning of the constraint will be evident from the relation obtained by equating r.h.s of (\ref{rn}) to logarithm of matrix model partition function in hamiltonian representation:

\be
\lan f;\tau =\infty| e^{-\int H_{\rm string} d\tau} O|i,\tau=-\infty\ran= {\rm ln} \lan f_M|  e^{\int_{t_o}^\infty dt H}e^{W(t_o)}e^{\int_{-\infty}^{t_o} dt H }|i_M\ran
\la{finalrelation}
\ee
Now note above, from r.h.s of (\ref {finalrelation}) that  the nonconservation of matrix model hamiltonian H, in the presence of macroscopic loop operator $W(t_o)$, in turn implies (in the l.h.s of (\ref{finalrelation}))  nonconservation of string hamiltonian $H_{\rm string}$ in the presence of boundary operator O.
\vskip1mm
However $H_{\rm string}$ decides mass shell condition of string \cite{gsw} and nonconservation of  $H_{\rm string}$ impliess that string is going to its off shell in the presence of boundary operator O, which is never feasible.   
\vskip1mm
Hence our claim that matrix model hamiltonian must be conserved in the presence of D brane operator is justified and consequently the constraint we impose (\ref{initialcondition}) is justified provided the relation (\ref{leak}) is true, which we are going to see in section 2.2. 
           
Note that, when inserted in the path integral, the above condition will act as an operator constraint on  the operator ${e^{\int dt W(t)\d(t-{t_o})} }$.  This condition was yet unrecognized in matrix model. The origin of this condition will be discussed in details in section 2.2. 
\vskip2mm
So we are going to show that the matrix model parttion function in the presence of a D brane is integrable if and only if the above condition(\ref{initialcondition}) is applied.  Here we consider Type 0A MQM with simplest macroscopic loop operator, which is the operator in NS sector(as prescribed in \cite{taka}), which we make stationary (\ref{initialcondition}) in time direction, exactly at the point of insertion, i.e at $t=t_o$, and with it we show that partition function indeed have an integrable structure.    We obtain the string equation. 
 \vskip2mm

The plan of this work is that in section 2.1 we are going to consider basic Type 0A MQM in the presence of D brane, with uncompactified time and review the theory.
In section 2.2 we are going to explain that, why we need to introduce the condition(\ref{initialcondition})in the matrix model path integral in the presence of D brane, in order to avoid nonconservation of MQM hamiltonian. In section 2.3, in order to  state the string theoretical origin of this operator constraint to the matrix model in the presence of D brane, first we review the existing results.  In section 2.4, we discuss the string theoretical interpretation of the constraint (\ref{initialcondition}).  In section 3 we consider MQM compactified on a circle in the presence of D brane. From path integral approach we are going to show that the partition function can be expressed as Fredholm determinant if and only if the condition (\ref{initialcondition}) is applied.  We have explicitly evaluated the thermal partition function and have shown that without application of this  condition  the partition function diverge. In section 4 we have discussed scattering in semiclassical regime. In section 5 we have considered string theory in the presence of momentum modes and have shown that the partition function in this background have an integrable structure if we apply the condition (\ref{initialcondition}).

 \section{Type 0A  MQM in the presence of the D-brane with the operator constraint}
\setcounter{equation}{0}
\subsection{The review of the theory of Type 0A  MQM in the presence of D brane }
Let us start with the MQM of type 0A theory in two dimensions, which is the decoupled world volume theory of (stable) D0 \textendash brane  and
anti D0\textendash branes.  A spacelike D0 \textendash $\overline{D0}$ pair, i.e with Neumann boundary condition in Liouville direction and Dirichlet in matter direction, gives a macroscopic loop observable of the matrix model after Laplace transformation \cite{newhat}.  We are going to consider Type 0A MQM in the presence  of this operator and study the relevant physics.  Here is a brief review of the Type 0A MQM. In the background with no net D0-brane charges, the matrix model 
has $U(N)\times U(N)$ gauge symmetry.  This is the
case we are going to consider.  We have the $U(N)\times U(N)$ gauge field $A_0$ and bifundamental tachyon $\Phi$, 
\begin{equation}  
{A_o}=\left(
\begin{array}{cc}
A & 0  \\
0 & \tilde{A}  \\
 \end{array}
\right),
\la{a}
\end{equation}
\begin{equation}  
{\Phi}=\left(
\begin{array}{cc}
0 & M  \\
{M^\dag} & 0 \\
 \end{array}
\right).
\la{m}
\end{equation}
The action is 
\be
\int dt Tr\left[ {({D_o}M)}^\dag {{D_o}M} +{\f{1}{2\alpha^\prime}}{M^\dag}M \right],
\la{matrixaction}
\ee
where 
\begin{equation}
{D_o}M={\partial_o}M +iAM-iM{\overline{A}}.
\label{derivative}
\end{equation}
As M is a complex matrix so we  denote M by Z, ${M^\dag}=\CZ$
\ber
{D_o}Z={\partial_o}Z+iAZ-iZ{\tilde{A}}  
&\quad;\quad& {({D_o}M)}^\dag={{\overline{D}}_o}{\CZ} ={\partial_o}\CZ  +i{\tilde{A}}\CZ - i\CZ A.
\label{derivative}
\eer
Now define
\begin{equation}
{{Z}_{\pm}}={\al} Z \pm  {D_o}Z \quad ;\quad   {{\CZ}_{\pm}}={\al}\CZ \pm  {\overline{D}_o}\CZ. 
\label{lightcone1}
\end{equation}
The Type 0A matrix model action in terms of the light cone variable
\be
S=\int dt Tr\left[ {\CP} {D_{A}} {\ZM} + \ZP \overline{({D_{A}} Z )}+{\frac{1}{2}}(\CM\ZP +\CP\ZM)\right].
\la{lagrangian}
\ee
The gauge field A acts as a lagrange multiplier which projects the theory onto singlet wave functions.  Its shown in \cite{yin,newhat} that type 0A MQM when projected to singlet sector can be represented by non-relativistic free fermions
moving in a two dimensional upside-down harmonic oscillator
potential
\be
 \hat H = {\f{1}{2}}(\hat{p}_x^2+\hat{p}_y^2)-{\f{1}{4\alpha^\prime}} (\hat{x}^2+ \hat{y}^2).  
\la{hamil}
\ee
The theory has different independent sectors labeled by net D0-brane charge q, which is the same as the angular momentum 
$\hat J = \hat x \hat p_y-\hat y\hat p_x$ \cite{newhat}.  Here we will consider the case where there is no net D0 \textendash brane charge, namely the $J=0$
sector. Now with $z,{\overline{z}} = x\pm iy; $ and light cone variable are as defined in (\ref{lightcone1} ) we have the hamiltonian
\begin{eqnarray}
{H_o}
&=& -{\frac12}({\hat{z}_+}{\hat{\overline{z}}_-}+{\hat{\overline{z}}_+}{\hat{z}_-}- {\frac{2i}{\sqrt{2{\alpha^\prime}}}}) \nonumber \\
&=& {\mp} {\frac{i}{\sqrt{2{\alpha^\prime}}}}\lbrack{z_{\pm}}{\frac{\partial}{\partial {z_{\pm}}}}+
{\overline{z}_{\pm}}{\frac{\partial}{\partial {\overline{z}_{\pm}}}}+1\rbrack.
\label{hamiltonian}
\end{eqnarray}
The commutation relation satisfied by these operators
\begin{eqnarray}
& & \lbrack\hzp,\hcm \rbrack =\lbrack \hcp,\hzm \rbrack=2 {\frac{i}{\sqrt{2{\alpha^\prime}}}},\nonumber\\
& & \lbrack\hzp,\hzm\rbrack =\lbrack \hcp , \hcm\rbrack =0,
\label{commutator}
\end{eqnarray}
so that
\begin{equation}
{\hat{\overline{z}}_+}=-{\frac{\partial}{\partial{z_-}}}  {\quad\quad;\quad\quad}{\hat{z}_-}={\frac{\partial}{\partial{\overline{z}_+}}}.
\label{relation1}
\end{equation}
We have Schrodinger equation 
\begin{eqnarray}
i {\frac{\p}{\p t}} {\Psi}(\cp,\zp,t)
&=&  {\mp}{\frac{i}{\sqrt{2{\alpha^\prime}}}}
 \lbr(\zp {\frac{\p}{\p\zp}}+ \cp{\frac{\p}{\p\cp}}+1\rbr {\Psi}(\zp,\cp,t).
\label{schrodinger}
\end{eqnarray}
Note, here we have absorbed the Vandermonde determinant in the wave function so that the  wave function ${\psi}$ in   (\ref{schrodinger}) describes a fermion.
\vskip3mm

Now consider the type 0A matrix model in the presence of  D brane which arises when we insert an operator ${e^{\int dt W(t)\d(t-{t_o})} }$ in the matrix model path integral where $W(t)$ is the Laplace transform of the macroscopic loop operator(\cite{taka}, \cite{openclosedual}).  In the dual two dimensional type 0A theory this means that there is one Euclidean $ D0\textendash \overline{D0}$ brane is localized at time $t_0$.  The branes extends along the Liouville direction after the Laplace transformation.  The macroscopic operators can be divided into NSNS and RR sector
part such that they correspond to the NSNS and RR sector part of
the D-brane boundary state.  Moreover,since we know that there are two types of (FZZT-like) boundary states
$|B(\epsilon)\ran$ according to the spin structures there should
be two macroscopic operators $W^{(\epsilon)}$ with $\epsilon=\pm$ in each sector.  First consider the simplest expression of macroscopic loop operator which is the one in NS NS  sector as prescribed  in \cite{taka} and expressed as  ${e^{-l{M^\dag}M({t_o})}}$.  Now, consider the Laplace transform of the operator  
\begin{eqnarray}
\int {\frac{dl}{l}} e^{-{\mu_B^2}l}{e^{-l{M^\dag}M}}
&=& -Tr\log (1+{\frac{{M^\dag}M}{\mu_B^2}}) \\
&=&  -Tr\log (1+{\frac { \CZ Z}{\mu_B^2}})\\
&=& -\sum\log (1+{\frac{\cz z}{\mu_B^2}})\nonumber\\
&=& -\sum\log (1+{\frac{ ({{z }_+}+{{z}_-} )({{{\overline{z}}}_+}+{{{\overline{z}}}_-} )  }{\mu_B^2}})\nonumber\\ 
&=& -\sum\log (1+{\frac{ {{{\overline{z}}}_+}{{z}_+}+ {{{\overline{z}}}_-}{{z}_-}+ {{{\overline{z}}}_+}
{{z}_-}+ {{{\overline{z}}}_-}{{z}_+} }{\mu_B^2}})\n
&=& W(\cp,\zp,\cm,\zm).
\label{macr}
\end{eqnarray}
(Here $\sum$ implies sum over the eigenvalues ).  Now the macroscopic loop operator for ${{NS}^-}$ sector can be expressed as 
\begin{eqnarray}
W
&=& -\sum\log (1+{\frac{ {{{\overline{z}}}_+}{{z}_+}+ {{{\overline{z}}}_-}{{z}_-}+ {{{\overline{z}}}_+} 
{{z}_-}+ {{{\overline{z}}}_-}{{z}_+} }{\mu_B^2}})\nonumber\\
&=&  -\sum \lbrace{\displaystyle\sum_{n=1}^{\infty}}{\frac{{(-1)}^n}{n}}{{\lbrack{\frac{ {{{\overline{z}}}_+}{{z}_+}+ {{{\overline{z}}}_-}{{z}_-}+ {{{\overline{z}}}_+}{{z}_-}+ {{{\overline{z}}}_-}{{z}_+} }{\mu_B^2}}\rbrack}^n}\rbrace.
\label{macrons}
\end{eqnarray}

\subsection{The origin of the operator constraint to the matrix model in the presense of D brane}
 
Here we are going to state the origin of the condition (\ref{initialcondition}), as we introduced in the introduction(section 1.2), from general consideration as well as from Ward identity. 
Note, from the analysis of previous section 2.1, the path integral  over the Euclidean time in the presence of D brane is expressed as 
\be
\int\prod{d\ZP}d{\ZM}d{\CP}d{\CM}dA d{\tilde{A}} {e^{-\int dt \lbr\b L(\ZP,\CP,\ZM,\CM,A,{\tilde{A}} ) - W\d(t-{t_o})\rbr}}.
\la{patitingal}
\ee
Since,path integral,implies time evolution w.r.t hamiltonian operator $H_o$, from an initial state at $t\rightarrow-\infty$ to the state at $t\rightarrow \infty$, where $H_o$ is the hamiltonian of type 0A MQM, so the insertion of the D brane-operator ${e^{\int dt W(t)\d(t-{t_o})} }$ in the path integral  apparently implies a shift\footnote{Note that as $W(\cp,\zp,\cm,\zm)$ in any sector  expressed in light cone variable, so does not involve any derivative.  Hence we can just add it to hamiltonian or lagrangian as a potential localized in $t_o$} \footnote{Note, when we are adding the term $ W\d(t-{t_o})$
in the expression of the hamiltonian from the operator ${e^{W({t_o})}}$ it is supposed to add in the hamiltonian the terms like 
$\lbr \b\int_{{t_o}-\epsilon}^{{t_o}+\epsilon} dt H,\int_{{t_o}-\epsilon}^{{t_o}+\epsilon} d{t^\prime}
 W\d({t^\prime}-{t_o})\rbr$ + ....higher commutators $= e^{-W(t_o)}\lbr\, \b\int_{{t_o}-\epsilon}^{{t_o}+\epsilon} dt H 
 \,\rbr e^{W(t_o)} +...O({\b \epsilon})^2..\,$.  However as around the ${\d-{\rm function}}$, $\epsilon$ can be made arbitrarily small i.e $\epsilon < < {\f{1}{\b}}$, so upto quasiclassical limit one can just  put these terms to zero while in the classical limit these terms are trivially zero.}

in the  hamiltonian
\be  
\b H_o \rightarrow \b H_o + W\d(t-{t_o}),
\la{shift}
\ee

( in Euclidean time). Note here we are going to consider  complete  quantum theory along with the macroscopic loop operator W.  
 However before proceeding  note that the operator $e^{\int dt W\d(t-{t_o})}$  which is localized at $t={t_o}$, in general  breaks the time translation symmetry of effective MQM action
\be  
\int dt \b L \rightarrow \int dt [ \b L - W\d(t-{t_o})],
\la{shiftlagrangian}
\ee
(as follows from (\ref{shift}))  because $W\d(t-{t_o})$ is localized in time.   So as far as  matrix model action in the presence of brane is concerned  (as considered in \cite{taka}) this is describing nonconservation of  MQM hamiltonian exactly at  $t_o$.
\vskip1mm
The above argument about nonconservation of MQM hamiltonian in the presence of D brane operator will become clear from the consideration of Ward identity.
\vskip1mm
  Consider MQM path integral in the presence of an operator $e^{W(t_o)}$, for which under any infinitesimal variation in time $t\rightarrow t+\epsilon(t)$, Ward identity implies \footnote{Here we have used the fact that equation of motion  in the presence of an insertion $e^{W(t_o)}$ is satisfied}
 
\ber
\d(t- {t_o})\d\lan  e^{W(t_o)}\ran 
&     =       & -  \p_t \lan{ H_o} (t) \,e^{W(t_o)}\ran\n
& \Rightarrow & \lan \, \d W({t_o})\, e^{W(t_o)} \,\ran + \lim_{\epsilon\rightarrow 0} \int_{t_o - \epsilon}^{t_o +\epsilon} \p_t \lan{ H_o} (t)  \,   e^{W(t_o)}\,\ran = 0,
\la{ward}
\eer
 
where $\d W$ is the variation of W due to infinitesimal time translation at fixed time ${t_o}$.

 The second line in (\ref{ward}) is coming by integrating the first one  over an infinitesimally small interval ${t_o} -\epsilon$ to ${t_o} +\epsilon$. Now since this is true for any $t_o$ in general, the  time translation invariance implies(\ref{ward}) 
 \be 
 \int_{t_o - \epsilon}^{t_o +\epsilon} dt\, \p_t \lan\,{ H_o} (t)\,\ran + \d \lan \,W({t_o}) \,\ran = 0. 
 \la{leak}
\ee
Now we have seen in section 1.2  that $ \int_{t_o - \epsilon}^{t_o +\epsilon} dt\, \p_t \lan\,{ H_o} (t)\,\ran $ must be zero or in other words matrix model hamiltonian must be conserved in the presence of D brane operator. So (\ref{leak}) implies
     
\be
\delta\lan  e^{\int dt W({t})\delta(t-{t_o})}\ran =\lan \delta e^{\int dt W({t})\delta(t-{t_o})}\ran = 0.  
\la{impose}
\ee

 In   subsection 2.4 we will show that the consequence of this condition   are in exact agreement with that of  string theory.  To understand  the impact of this condition  in  MQM first we need to  write down the Schrodinger equation and study the Hilbert space.
We have the time dependent Schrodinger equation for a single fermion \footnote{When we consider the insertion of $\b$ factor, for the macroscopic loop operator we have the expression  ${\f{1}{\b}}{\hat{W}}$, however when we rewrite the Schrodinger equation in terms of the eigenvalues x,y which corresponds to the real and imaginary part of eigenvalue z we have the Schrodinger equation
\\
$$\lbr{\f{1}{\b^2}}{\f{\p^2}{\p{x^2}}} + {x^2} + {\f{1}{\b^2}}{\f{\p^2}{\p{y^2}}} +  {y^2} + {\f{1}{\b}}{\hat{W}}\delta(t-{t_o}) \rbr\psi =E\psi$$
In the double scaling limit we take $x,y \rightarrow {\sqrt{\b}}x, {\sqrt{\b}}y$ \cite{KIR}, which gives the Schrodinger equation  (\ref{schrodinger} )}  in Minkowskian time, as implied from (\ref{shift}),
\begin{eqnarray}
\lbrack i {\frac{\partial}{{\partial} t}} 
&-& i{\delta}(t-{t_0})W(\cp,\zp,\cm,\zm)
\rbrack{\Psi}(\cpm\zpm,t)\nonumber\\
&=&  {\mp}{\frac{i}{\sqrt{2{\alpha^\prime}}}}
 \lbr(\zp {\frac{\p}{\p \zpm }}+\cp {\f{\p}{\p\cpm}}+1{\rbrack}{\Psi}(\zpm,\cpm,t).
\label{schrodingerw}
\end{eqnarray}
For t away from ${t_o}$ we have time independent Schrodinger equation
\begin{eqnarray}
 i {\f{\p}{\p t}} {\psi}(\cpm\zpm,t)
&=& {\mp}{\f{i}{\sqrt{2{\alpha^\prime}}}}
 \lbrack(\zpm{\f{\p}{\p \zpm}}+\cpm {\f{\p}{\p \cpm}}+1 \rbrack {\Psi}(\cpm\zpm,t)= E{\psi}(\cpm\zpm,t).
\label{schrod}
\end{eqnarray}
The free fermion solution with energy E is
\begin{equation}
{\psi^E_{o \pm}}({z_\pm},t) ={e^{-iEt}}{e^{\mp i{\f {\phi_o (E)}{2}} }}{{(\cpm\zpm)}^{{\pm}iE-{\frac{1}{2}} }},
\label{freesolution}
\end{equation}
where we have chosen ${\alpha^\prime}=2$ and ${\phi_o}(E) $ is determined from biorthogonal property (discussed in the Appendix) and given by 
\be
e^{i{\phi_0(E)}}
= {\Gamma(iE+{\f{1}{2}})\over \Gamma(-iE+{\f{1}{2}})}. 
\la{explicitform}
\ee
Now consider the commutation relation
\begin{eqnarray}
\lbrack {H_o},{\hat{z}_+}\rbrack
&=& -i{\quad\quad;\quad\quad}\lbrack {H_o},{\hat{\overline{z}}_+}\rbrack=-i, \nonumber \\
\lbrack {H_o},\hzm\rbrack
&=& \quad i{\quad\quad;\quad\quad}\lbrack {H_o},{\hat{\overline{z}}_-}\rbrack=i. 
\label{com}
\end{eqnarray}
This implies $\hcp\hzp$ and $\hcm\hzm$ when acts on a state $|E\ran$ expressed as $\hcp\hzp|E\ran= |E-i\ran$ and $\hcm\hzm|E\ran= |E+i\ran$.  These states can be represented as $|E\pm ni\ran$. These describe  a different Hilbert space \cite{thesisalexandrov} which can be understood as their inner product with the state $|E\ran$ either diverge or zero.  These states actually can be identified with the discrete tachyonic states  over the matrix model ground state \cite{lectjevicki,thesisalexandrov}. 
\vskip.1mm
Now let us come to the meaning of the constraint(\ref{impose}).  First we will find the expression for the constraint and then solve it in the context of the matrix model path  integral for the macroscopic loop operator we considered or in the more complicated case to reach to the right expression  free energy.  Consider the v.e.v of the operator in single fermionic state which is, following (\ref{hamiltonian},\ref{macrons}) given by 
 \be
\lan e^{W({t_o})}\ran = \lan e^{{\rm log} (1+{ \f{\cp\zp+\cm\zm-2{H_o}}{\mu_B^2}})}\ran = \lan (1+{ \f{\cp\zp+\cm\zm-2{H_o}}{\mu_B^2}})\ran.
\la{vev}
\ee
(Here we have shown the expectation value w.r.t the single fermionic state. This we could do because the theory is projected singlet sector and N fermionic state is just the direct product of each).  So (\ref{impose}) along with (\ref{com}) gives the constraint

\be
 \lan e^{W({t_o})} \d W({t_o})\ran = 0 \quad\quad\Rightarrow \quad\quad \lan e^{W({t_o})} \lbr \cp\zp(t_o)-\cm\zm(t_o)\rbr\ran = 0.
\la{ve}
\ee
which implies the constraint
\be
\lan \cp\zp(t_o)-\cm\zm(t_o)\ran = 0.
\la{imposedcondition}
\ee

So, with the above constraint we can write the path integral with the D-brane-operator inserted as

\ber
\int d\cp d\zp d\cm d\zm {e^{-\int_{-\infty}^{\infty } dt\lbr \b L -W( \cp,\zp,\cm,\zm;{t_o})\rbr}} 
&=& \int \displaystyle\prod_{t\le{t_o}}d\cp d\zp d\cm d\zm {e^{-\int_{t_o}^{\infty} dt L }}\n
\int d\cp(t_o) d\zp(t_o) d\cm (t_o)d\zm(t_o)
& & \d(\cp({t_o})-\cm({t_o}))\d(\zp({t_o})-\zm({t_o}))
 e^{W(\cp\zp,\cm\zm, H_o ; \, {t_o})}\n
 \int \displaystyle\prod_{t\ge{t_o}}d\cp d\zp d\cm d\zm {e^{-\int_{-\infty}^{t_o} dt L }}
&=& \int d\cp d\zp d\cm d\zm {e^{-\int_{-\infty}^{\infty } dt\lbr \b L -W(\cpm\zpm(t_o),H_o)\rbr}}.
\la{formal}
\eer
Its important to note that the effect of the constraint is only to project W at its physical sector without imposing any boundary condition to original lagrangian which happens due to $\d (t- t_o)$ factor as we explained.  Projection implies we can get same path integral expression or the transition amplitude by expressing  W either in $\cp\zp$ or in $\cm\zm$ mode which happens due to the fact that wave function associated with the either mode appears to be  same at $t_o$.

\subsection{String theoretical interpretation:Review of formal result}

In order to show that the constraint we imposed(\ref{impose}), leads to the right matter wave function associated with the boundary state in the collective field theory,  first we need to review the boundary state formalism in 2D theory and some collective field theory. The boundary state of 2D spacelike brane is expressed as 
$ |B_{\rm (Super FZZT)}(\mu_B)\ran_{\phi}\otimes |D\ran_{X^0}$ where  $|D\ran_{X^0}$  for NS NS sector \cite{cardystate} is expressed as
$\CN \int_{-\infty}^{\infty} dP e^{iP{X^o}}|P\ran$  ( $\CN$ is the normalization factor) which in terms of the vertex operator can be written as 

\be
|B{\ran_X}=\CN \int_{-\infty}^{\infty} d{P_m} \lbr e^{i{P_m}(X-{X^o})} \rbr  |0\ran + {\rm descendants}.
\la{bstate}
\ee
The string endpoints are localized at $X= X^o$ and the matter part of the boundary state form the representation of $\delta(\hat{X }- {X^o})|0\ran$. 
Boundary state of Super Liouville  NS NS sector is given by
\be
|B{\ran_L} = \CN_l {\int_0^\infty} dPU(P_l)|P_l\ran \quad;\quad U(P_l)= {\f{\pi {\rm cos}2\pi s {P_l} }{{\rm sin }h (\pi {P_l} )}} \quad;\quad |P_l\ran = (1+ {\f{L_{-1} \tilde{L}_{-1}}{P_l^2 +M^2}} + .....)|v_{P_l}\ran
\la{up}
\ee 
where $|v_{P_l}\ran$ primary macroscopic state associated with a vertex $e^{({\f{Q}{2}} + i{P_l})\phi}$, M is the mass of intermediate propagating mode. 
\\
The boundary state is the direct product of the matter  and Liouville part along with the ghost factors. Setting ${P_m} = {P_l} =P$ we have the primary part of the boundary state without ghost excitation mode, which is the superposition of the tachyonic field can be expressed in terms of the  operators from the state operator mapping as
\be
\int dP U(P)\lbrack e^{{iP(X-{X^o})}} + e^{{-iP(X-{X^o})}}\rbrack {e^{({\f{Q}{2}} +iP)\phi}}.
\la{mixbstate}
\ee
  Now in the 2nd quantized matrix model we can express the  macroscopic loop operator as 
\be
W(l,t)= \int  e^{l {\overline{z}{z}}}  {\psi^\dag}{\psi}= \int d \tau e^{l{\rm cosh}^2 \tau}
{\p _\tau}\eta (l,t),
\la{wmacro}
\ee
 where $\tau$ is the time of flight coordinate obtained from reparameterization ${\overline{z}{z}}={r^2}\quad;\quad {r^2} \sim {\rm cos}^2 {h\tau}$ , ${\psi^\dag}
{\psi} \sim {\p \tau}\eta (t,\tau)$ and $\eta$ is the massless bosonic field $ ({\p_t^2}- {\p_\tau^2})\eta(t,\tau) = 0$ which corresponds to the tachyon in the string theory at asymptotic $\tau$\cite{stringfield,GM,lectjevicki}.  $\eta$ corresponds to the fluctuation of the collective field $\phi ={\phi_o} + {\p_\tau}\eta $. $\eta$ satisfies Dirichlet boundary condition in $\tau$ direction \cite{loopstofields}, \cite{lectjevicki}. However note that when  associated with the Laplace transformed  macroscopic loop operator operator W which correspond to D brane boundary state, $\eta$ is no longer an ordinary state  but it corresponds to an Ishibashi state.  However, so far in the literature, the matter one point function of this Ishibashi state is not determined.  More precisely, consider the expression 
\be 
\eta(\tau,t) = \int_{-\infty}^{\infty} {\f{dp}{p}}{\tilde{\eta}}(p)(a(p) e^{-ipt}+b(p) e^{ipt}) {\rm sin}(p\tau).
\la{eta}
\ee
In the above, a(p) and b(p) are not determined.  In other words, given the fact that the boundary state (\ref{mixbstate}) is the Laplace transform of W in (\ref{wmacro}), we should be able to write both Liouville and matter one point funtion (i.e U(p) and $e^{iPX^O}$), in collective field theory, with $\eta(t,\tau)$ as the Ishibashi state. However in the existing literature, while the Liouville wave function $U(p)$ is well determined in matrix model, the matter wave function ($e^{iPX^O}\equiv e^{ipt_o}$) is not being expressed, i.e we do not have any information about a(p) and b(p) in (\ref{eta}).   In the next subsection we will show that when we impose the constraint (\ref{impose}) in collective field theory, we able to determine a(p) and b(p), so that W in (\ref{wmacro}) can be expressed in the form of (\ref{mixbstate}), with $\eta$ as Ishibashi state.      
\vskip2mm
\subsection{String theoretical interpretation : Imposition of the constraint} 
\vskip2mm

In this subsectioon, we will impose the constraint (\ref{impose}) to the collective field theory, and show, this will lead to the correct matter wave function for the boundary state in collective field theory.

In order to find the exact t   dependence consider the fact that the implication of (\ref{impose}) in the context of collective field theory is
\be
{\d_t}{\phi} {|_{t=t_o}} = 0, 
\la{imposend}
\ee
where as before $\d_t$ implies variation of the collective field $\phi$ due to infinitesimal variation in t, at fixed t.

 So (\ref{imposend}) implies  in (\ref{eta}) we have
\be
a= e^{ipt_o} \quad;\quad  b= e^{-ipt_o}
\la{ab}
\ee
So (\ref{eta}) is expressed as $\eta(\tau,t) = \int_{-\infty}^{\infty} {\f{dp}{p}}{\tilde{\eta}}(p)(e^{-ip(t-{t_o})}+e^{ip (t-{t_o})}){\rm sin}p\tau$.  Now integrating over $\tau$ we have $W(p,t)={\f{e^{-l\mu}}{2}} p{K_{ip}}(\mu l){\tilde{\eta}}(p)(e^{-ip(t-{t_o})}+e^{ip(t-{t_o})})$ where ${K_{ip}}(\mu l)$ is the Bessel's function which is the macroscopic wave function satisfying WdW equation \cite{taka}. This on Laplace transform $\int {\f{dl}{l}}  e^{-\mu_B^2 l}$ can be expressed as 
\be
W(p,t)=  U(p) \tilde{\eta} (p,t) \quad;\quad   U(p)= {\f{\pi {\rm cos}2\pi sp}{{\rm sin }h (\pi p)}}
\la{wp}
\ee
where we have
\be
 \mu_B^2=2\sinh^2(\pi s)|\mu| \ \ \
(\epsilon\cdot {\rm sign}(\mu)<0),\ \ \ \ \ \mu_B^2=2\cosh^2(\pi s)|\mu| \ \
\ (\epsilon\cdot{\rm sign}(\mu)>0).
\la{mub}
\ee
Note (\ref{impose}) and (\ref{imposend})are just the same constraint expressed in the 1rst and 2nd quantized formalism.
Given the form of $\eta$, on Laplace transform of (\ref{wmacro}) and on inverse Fourier transform of (\ref{wp})  we can express W as
\be 
W_p (\tau,t)= U(p){\tilde{\eta}}(p)\lbr e^{ip(t-{t_o})}+e^{-ip(t-{t_o})}\rbr=  U(p)\lbr  {\p_{\tau}}{\eta_p} (\tau+(t-{t_o}))+  {\p_{\tau}}{\eta _p}(\tau-(t-{t_o})) \rbr
\la{final}
\ee
where the suffix p implies pth component. We know $\eta$ corresponds to the space time tachyonic field in the asymptotic $\tau$ region which in the string theory side describes asymptotics of Liouville field $\phi$ which describes vanishing 	Liouville wall.  So we see that each component with momentum p in the expansion of W  describes the tachyonic operator ${\tilde{\eta}}(p)$ dressed with Liouville and matter wave function (\ref{bstate},\ref{up}) where W is symmetric under $P\rightarrow -P$.  So these are just in one to one correspondence with the state obtained in the expansion of the boundary state in (\ref{mixbstate}).  So (\ref{final}) is just the same as (\ref{mixbstate}) when we impose the constraint(\ref{imposend}). 
\\
So we see that imposition of the constraint(\ref{imposend}) leads to the right form of the matter wave function in the context of collective field theory.

\section{Type 0A MQM on a circle in the presence of D brane}
\setcounter{equation}{0}
\subsection{Evaluation of the free energy of Type 0A matrix model on a circle : A review}
We consider Type 0A matrix model compactified on a circle of radius R.  As considered in the previous section, there is no net background D0 brane charge and hence it is described with $ U(N) \times U(N) $ gauge symmetry. The partition function  in terms of the light cone variables is given by
\ber
 \int d{\ZP}d{\ZM}d{\CP}d{\CM}dA d{\tilde{A}} e^{{-\b\int_0^{2{\pi}R}} dt
 Tr\left[ {\CP {D_A} }{\ZM} + {\ZP \overline{({D_{A}} \ZM)}} + {\frac{1}{2}}(\CM\ZP +\CP\ZM)\right]},
\la{partition1}
\eer
where ${D_{A}}Z = {\partial_t}Z +  i\lbr {A}Z -Z{\tilde{A} }\rbr $ and A is as given in (\ref{a}).  Now we fix the gauge ${\partial_t}A =0 $, which sets $A$ and $\tilde{A}$ to their zero modes $A^{(0)}\equiv X/2\pi\alpha'$ and $\tilde{A}^{(0)}\equiv \tilde{ X}/2\pi\alpha'$, where in the T-dual theory X and ${\tilde{X}}$ corresponds to collective coordinate of D0 and anti D0 brane \cite{newhat}.  As before the gauge fixing
introduces the FP determinant \cite{yin} 
\be
\int db\,dc\exp({\rm Tr}b \partial_t D_t c)={ \prod_{i<j}}{\left( {\f  {\sin [(x_i-x_j)R/2] }  {(x_i-x_j)R/2} } \right)}^2 
{\left({\f{\sin [ (\tilde {x}_i-\tilde {x}_j)R/2]} {(\tilde{x}_i-\tilde {x}_j)R/2}}\right)}^2 ,
\la{fa}
\ee
where $x_i$ and $\tilde{x}_i$ are the eigenvalues of $X$ and $\tilde{X}$ respectively. Now the denominator gets cancelled  with the respective Vandermonde determinant so that the  usual measure factor $\Delta(x)^2\Delta(\tilde x)^2$ is converted to the measure for unitary matrices 
\be
{ \prod_{i<j} \sin^2({(x_i-x_j)R\over 2})
\sin^2({(\tilde x_i-\tilde x_j)R\over 2}) } .
\la{me}
\ee 
Note these are the measure for unitary matrices $U=e^{\f{iXR}{2}},~ \tilde U = e^{ \f{i\tilde XR}{2}}$.which are holonomy factors (we chose ${\alpha^\prime = 2}$). Therefore the natural variables to be integrated over are the ``holonomies" $U=e^{2i\pi A^{(0)}R},~ \tilde U = e^{2\pi i\tilde{A}^{(0)}R}$.  Once we gauge fix A and $\tilde{A}$  The partition function  depends on the gauge field only  through the global holonomy factor,  given by the unitary matrix
\be
\Omega =
\hat T e^{i\int_0^{2\pi  R} A(t)dt} \quad ; \quad \tilde{\Omega}=\hat T e^{i\int_0^{2\pi  R} \tilde{A}(t)dt}. 
\la{omega}
\ee
 In the $A=const$ gauge, in the path integral, the constant modes of A can be absorbed by redefining the fields ${\ZM},{\CM}$ as
\be
\ZM(t) \rightarrow {e^{-iAt} \ZM(t)e^{i\tilde{A}t} } \quad ; \quad \CM(t) \rightarrow {e^{-i\tilde{A}t} \CM(t)e^{iAt} },
\la{redefinition}
\ee
which replaces the periodic boundary condition  $$\ZPM(2{\pi}R)=\ZPM(0)\quad;\quad  \CPM(2{\pi}R)=\CPM(0) $$ by a $SU(N)$-twisted one \cite{intflow},\cite{holography}
\ber
\ZP(2{\pi}R)=\ZPM(0)  
&  & \quad;\quad \CP(2{\pi}R)=\CP(0) \n
\ZM(2{\pi}R)={\O}\ZM(0) {\tilde{\O}^{-1}} 
&  & \quad;\quad \CM(2{\pi}R)={\tilde{\O}}\CM(0) {\O^{-1}}, 
\la{twistedbc}
\eer
So in the constant A gauge  integration  with respect to the fields $\ZPM(x),\CPM(x)$  is Gaussian with  the determinant of the quadratic form equal to one.
Therefore it is reduced to the integral with respect to the initial values $\ZPM,\CPM=\ZPM,\CPM(0)$ of the  action evaluated  along the classical trajectories, which satisfy the twisted periodic boundary condition (\ref{twistedbc}).  Therefore  the canonical partition function of the matrix model can be reformulated as an ordinary matrix integral with respect to the  hermitian matrices  $\ZP$ , $\ZM$ ; $\CP$ , $\CM$ and the unitary matrices 
$\O ,\tilde{\O}$, :
\be
{\cal{Z}}_N= \int  d \ZP d\ZM d\CP d\CM d\O d{\tilde{\O}} e^{ i\b Tr (\CP\ZM+\ZP\CM  -  q \CM\Omega \ZP \tilde{\Omega}^{-1} -  q \ZM\tilde{\Omega }\CP 
\tilde{\Omega})} ,
\la{parti}
\ee
 where we denote
\be
q= e^{2i \pi R}.
\la{q}
\ee
Now note the above expression can be written as 
 \ber
\cal{Z}_N
&=&\int  d \ZP
d\ZM d\CP d\CM
  d\O d{\tilde{\O} }
 e^{ i\b{\rm Tr} (\CP\ZM+\ZP\CM  -  q \CP\Omega \ZM \tilde{\Omega}^{-1} -  q \ZP\tilde{\Omega }\CM {{\Omega}^{-1} })}\n
&=&\int  d \ZP
d\ZM d\CP d\CM
  d\O d{\tilde{\O} }
 e^{ i\b{\rm Tr} (\CP\ZM+\ZP\CM  -  q \CP(\Omega \ZM{\Omega}^{-1} )\Omega\tilde{\Omega}^{-1} -  q \ZP\tilde{\Omega }{\Omega}^{-1} ({\Omega }\CM {{\Omega}^{-1} })}\n
&=& \int  d \ZP
d\ZM d\CP d\CM
  d\O d{\tilde{\O} }
 e^{ i\b{\rm Tr} (\CP\ZM+\ZP\CM  -  q \Omega\tilde{\Omega}^{-1}\CP(\Omega \ZM{\Omega}^{-1} ) -  q \ZP\tilde{\Omega }{\Omega}^{-1} ({\Omega }\CM {{\Omega}^{-1} })}\n
&=& \int  d \ZP^\prime d\ZM d\CP^\prime d\CM
  d\O d{\tilde{\O} }
 e^{ i\b{\rm Tr} (\tilde{\Omega }{\Omega}^{-1} {\CP^\prime}\ZM+{\CM}{\ZP^\prime}\Omega \tilde{\Omega}^{-1}  -  q \CP^\prime\Omega \ZM{\Omega}^{-1}-  q \ZP^\prime {\Omega }\CM {{\Omega}^{-1} }})\n
&=& \int  d \ZP^\prime d\ZM d\CP^\prime d\CM
  d\O d{\O^\prime} 
 e^{ i\b{\rm Tr }(\O^\prime {\CP^\prime}\ZM+{\CM}{\ZP^\prime}{{\O^\prime}^{-1}} -  q \CP^\prime\Omega \ZM{\Omega}^{-1}-  q \ZP^\prime {\Omega }\CM {{\Omega}^{-1} }}),
\la{partim}
\eer
where we define $ \ZP\tilde{\Omega }{\Omega}^{-1} =\ZP^\prime\quad;\quad \Omega\tilde{\Omega}^{-1}\CP = {\CP^\prime}$ \quad;\quad${{\Omega^\prime}}= \tilde{\Omega}{\Omega}^{-1}$.The last expression implies replacing $\tilde{X}$ by $\tilde{X} -X$, both running over the infinite real line.  The redefinition of the variables will keep the measure invariant.  So by generalizing  Harishchandra-Itzykson-Zuber integral  we can write the above partition function as \footnote{$( \int dU e^{iTr UX{U^{-1}}Y} =  Const.{\frac{\det {e^{i{x_k}{y_l}}}}{\Delta(x)\Delta(y)}}$where ${x_k}$ and ${y_l}$ are eigenvalues of X and Y and ${\Delta(x)}$ ${\Delta(y)}$ are Vandermonde determinant given by 
 $\Delta(x)={\displaystyle\prod_{i\le j}}({x_k}-{x_l})$ }, \footnote{To express the part involving ${\Omega^\prime}$ we used the fact that 
 in the integral  $( \int dU e^{i{\rm Tr }UX{U^{-1}}DY}$ where D is a complex diagonal matrix with ${D^{-1}} ={D^\dag}$ and $Y=VyV^{-1}$ where y is the eigenvalue of Y and V is the unitary matrix diagonalizing Y. Now  we can write $DY = {V^\prime}d \,y{{V^\prime}^{-1}}$ for some other diagonalizing matrix ${V^\prime}$ which exploits the fact that ${\rm Diag}({{(DY )}^\dag}DY)={\rm Diag }({{ Y }^\dag}Y)={\rm Diag}({V}{y^*}y {V^\dag})$ 
 which implies the above expression (d is eigenvalue of D).So following the formal derivation of the integral we can write 
 
  $( \int dU e^{iTr UX{U^{-1}}DY}  = Const.{\frac{\det {e^{i{x_k}{d_{kl}}{y_l}}}}{\Delta(x)\Delta(y)}},$ which is nonzero only when $k=l$. for any diagonal matrix D.Note,we are not summing over k and l.Denominators gets canceled with the Vandermonde determinants appearing from $\CP, \ZP, \CM,\ZM $.}\cite{intflow}.
 
\ber
{\cal{Z}}_N(t)
&=& \int\limits _{-\infty}^\infty
  \prod_{k=1}^N
 [d \zp_k ][d \zm_k] [d \cp_k ][d \cm_k] [d {\O^\prime}_{kk}]\nonumber\\
& & \lbr \det_{jk}\left( e^{i {{\O^\prime}_{jk}}\cp_j \zm_{k}}\right) 
 {\rm det}_{jk} \left( e^{-iq \cp_{j} \zm_k}\right)
{\rm det}_{jk}\left( e^{i  {{\O^\prime}^{-1}}_{jk} \cm_j \zp_{k}}\right)\n
& & {\rm det}_{jk} \left( e^{-iq \zp_{j} {\cm_k}}\right)\rbr,
\la{paran}
\eer
where $ {\O^\prime_{jk}}$ which has only  diagonal elements nonzero.  Now we show that the  grand canonical partition function can be written as a Fredholm determinant
\be
Z(\mu,t )= {\rm Det}(1+e^{-\tpr\beta \mu} K _+ K_- ),
\la{fredholm}
\ee
where  
\ber
\lbr {K_+} f \rbr({\cm\zm}) 
&=& \int  [d\cp][d\zp] dt 
{e^{  i (t\cp \zm +{t^{-1}}\cm\zp)} }f({\cp\zp}) ,\n
\lbr {K_-} f \rbr ({\cp\zp}) 
&=& \int [d\cm][d\zm]   {e^{ -iq(\cp \zm +\cm\zp)} }f({\cm\zm}) .
\la{k}
\eer
\be
{K_+}{K_-}f(\cp\zp) =\int  [d\cm][d\zm] dt 
e^{ i (t\cp \zm +{t^{-1}}\cm\zp)}\int[d\cp^\prime][d\zp^\prime] e^{ -iq(\cp^\prime \zm +\cm\zp^\prime)} f({\cp^\prime}{\zp^\prime}).
\la{kk}
\ee
Note that  t and ${t^{-1}}$ denote the diagonal elements of ${\O^\prime}$ and ${{\O^\prime}^{-1}}$ corresponding to $\cpm, \zpm$.  Now note when we evaluate the determinant in a diagonalizable basis which is naturally given by  $f(\cpm\zpm)$, ${K_+}{K_-}f(\cpm\zpm)$ will be independent of t i.e $\int dt$ will come out as an overall factor.  So following the analysis of \cite{intflow} the grand canonical partition function $ \displaystyle\sum_N {e^{-\tpr\b N\mu} {\cal{Z}}_N }$  can be expressed as the Fredholm determinant  (\ref{fredholm}) which is  same as that of $c=1$ matrix model.
  Now  following \cite{stroncircle} we can express the partition function  as  $Tr exp\lbr-\tpr\b H\rbr$. The gauge field A project the theory to singlet sector so that in the absence of perturbation, the grand canonical partition function 
is given by the Fredholm determinant
\be
{\cal{Z}}(\mu) = {\rm Det}(1+ e^{-\tpr\b(\mu+H_0)}),
\la{pa}
\ee
which must be same as (\ref{fredholm}).  This can be interpreted as the grand canonical finite-temperature partition function  for a system of non-interacting fermions in the inverse Gaussian potential.  The Fredholm determinant can be computed once we know a complete set of eigenfunctions for the one-particle Hamiltonian ${H_o}$.  Now in order to evaluate the free energy  we need to  find the density of states, it is conventional to introduce a cutoff $\Lambda$.

There is  no momentum flow through the wall $\cz z ={x^2}+{y^2}=|{\Lambda}|^2$ is implied by the condition 
 $({\hat{x}}{\hat{p}_y} + {\hat{x}}{\hat{p}_y}){\psi_\pm}(x,y){|_{({x^2}+{y^2}= |{\Lambda}|^2 )}}=(\hcp\hzp-\hcm\hzm){\psi_\pm}({\overline{z}},z) {|_{ ({\overline{z}}z = |{\Lambda}|^2 )}} = 0 $, which has a solution
\be 
\psi ^E_+({\Lambda}) = 
\psi ^E_-({\Lambda}).
\la{boundary}
\ee
This condition  is satisfied for a discrete 
set of energies $E_n( n\in Z )$
defined by
\be
\phi_0(E_n) - E_n\log \Lambda +2\pi n=0.
\la{ph}
\ee
 From  (\ref{ph}) we can find the density of the energy levels in the 
confined system
 \be
\rho(E)= {\f{{\rm log} \Lambda}{ 2 \pi\b}} - {\f{1}{2\pi\b} } {\f{d\phi_0(E) } {d E}} ,
\la{density}
\ee

as derived in \cite{intflow}
Now we can calculate free energy
${\cal{F}}(\mu, R)= {\rm log}{\cal{Z}}(\mu, R)$ as
\be
\CF(\mu, R)=  \int_{-\infty}^\infty
d E\, \rho(E)\log\left[1+e^{-\tpr\b(\mu+E)}\right],
\la{free}
\ee
with the density (\ref{density}). Integrating  by parts  in  and dropping out the $\Lambda$-dependent piece,
we get
\be
{\cal{F}}(\mu, R) =-{1\over{ 2\pi\b}}\int d\phi_0 (E)
\log \left( 1+e^{-\tpr\beta(\mu+ E)}\right)=-
R \int_{-\infty}^{\infty}dE
{ \phi_0(E) \over 1+e^{\tpr\beta(\mu+ E)}} ..
\la{fred}
\ee
We close the contour of integration in the upper half plane and  take the integral as a sum of residues. This gives for the free energy
\be
{\cal{F}}= -i \sum _{r=n+{\f{1}{2}} > 0}
\phi_o\left( i r/R-\mu\right).
\la{res}
\ee
As the Fredholm determinant is similar to that of $c=1$ MQM so following the analysis of \cite{intflow}, \cite{yin} we can see 
From (\ref{res}) it follows that \cite{nonperturbative}
\be
2{\rm sin }{\frac{\p_\mu }{2 {\b R}} }\cdot \CF(\mu ) = \phi_o (-\mu) .
\la{phifrerelation}
\ee
Also its shown that the free energy can be expressed as
\be
\CF_{\rm pert}(\mu)_{\{t_k=0\} }
= - \frac{R}{2}\mu^2 \log \frac{\mu}{\Lambda} -
\frac{R + {1\over R}}{24} \log \frac{\mu}{\Lambda} +
R\sum\limits_{h=2}^\infty \mu^{2-2h} c_h(R),
\la{genusexpansion}
\ee

\subsection{ Evaluation of free energy of Type 0A matrix model on a circle with D brane with the imposition of the constraint}

In this section we will consider the type 0A matrix model path integral in the presence of a D brane and show the grand canonical partition function  can be expressed as the Fredholm determinant.  We consider the brane in the NS NS sector and show that how to generalize the analysis for the brane in any other sector. Consider the path integral in the presence of the macroscopic loop operator localized at ${t_o}$, which is the generalization of (\ref {partition1}). The classical action will remain periodic even in the presence of D brane so we can express the 
\ber
 \int d{\ZP}d{\ZM}d{\CP}d{\CM}dA d{\tilde{A}} e^{{-\b\int_0^{2{\pi}R}} dt
{\rm Tr}\left[ {\CP {D_A} }{\ZM} + \ZP {D_A} {\CM} + {\frac{1}{2}}(\CM\ZP +\CP\ZM)\right] +{\rm Tr}{W(t_o)}    } ,
\la{partin}
\eer
 The macroscopic loop operator depends on diagonal elements only,  so the partition function  (\ref{parti}) and finally (\ref{paran}) can be expressed as
\ber
{{\cal{Z}}_N}(t)=\int\limits _{-\infty}^\infty
\prod_{k=1}^N
& &[d \zp_k ][d \zm_k] [d \cp_k ][d \cm_k] [dt_k]
{\rm det}_{jk}\left( e^{i {t^{-1}_{jk}}\zm_j \cp_{k}}\right)
{\rm det}_{jk} {\left( e^{-iq \zm_{j} \cp_k}\right)}
{\rm det}_{jk}\left( e^{i{t_{jk}} \cm_j \zp_{k}}\right)\n
{\rm det}_{jk} {\left( e^{-iq \cm_{j} \zp_k}\right)}
& & exp\lbr \displaystyle\sum_i log (1+{\frac{ \cp_i\zp_i+ \cm_i\zm_i+ \cp_i\zm_i+ \cm_i\zp_i }{\mu_B^2}})\rbr,
\la{partim}
\eer
(where we have off-diagonal term is zero , also we are not summing over j,k)
\\
where $\displaystyle\sum_i$ is coming from Trace.   Again above can be expressed as
\ber
{\cal{Z}}_N(t)
&=& \int\limits _{-\infty}^\infty
 \prod_{k=1}^N
 [d \zp_k ][d \zm_k] [d \cp_k ][d \cm_k] 
{\rm det}_{jk}\left( e^{i{t^{-1}_{jk}} \zm_j \cp_{k} }\right)
{\rm det}_{jk} \left( e^{-iq \zm_{j} \cp_k}\right)
{\rm det}_{jk}\left( e^{i {t_{jk}}\cm_j \zp_{k}}\right)\nonumber\\ 
& & {\rm det}_{jk} \left( e^{-iq \cm_{j} \zp_k}\right)
 \prod_{r=1}^N (1+{\frac{ \cp_r\zp_r+ \cm_r\zm_r+ \cp_r\zm_r+ \cm_r\zp_r}{\mu_B^2}}).
\la{partima}
\eer
Now in order to write the above expression we have used the fact that the classical action is periodic even in the presence of the macroscopic loop operator, 
$Z(2\pi R) = Z(0) $. However at the quantum level there is a discontinuity of state $|\psi(\tpr-\epsilon)\ran \ne |\psi(0) +\epsilon\ran$. This causes the absence of the vortex d.o.f. 

\textbf{Imposition of constraints}
 
Note with the help of the constraint (\ref{impose},\ref{imposedcondition})  we can write $\cp_r\zp_r+ \cm_r\zm_r$ in (\ref{partima}) as $2 \cpm_r\zpm_r$ inside the path integral and one can evaluate the partition function easily when $f({\cpm\zpm})$ is the function, forms the representation of $K_+$ and $K_-$  (\ref{k}, \ref{kk}). 
The action of W on f is given by 

\be
{\hat{W}}f({\cpm\zpm}) = (1+{\frac{ 2\hcpm\hzpm + \hcp\hzm+ \hcm\hzp}{\mu_B^2}})f({\cpm\zpm}).
\la{opw}
\ee
 Now,  the operator ${\hat{W}}$  does not introduce any interaction between the fermions, so no off diagonal terms from W. Now  from (\ref{fredholm}) and  (\ref{partim}) we can write the grand canonical partition $\sum_ N {e^{-\tpr \b N}}{\cal {Z}}_N$ as 
\be
{\cal {Z}}(\mu) = {\rm det} (1+e^{-2\pi R \b \mu} W K) .
\la{partioww}
\ee
Now in order to evaluate (\ref{partioww}) following (\ref{kk}) a representation of K is formed by the basis $f({\cp_i}{\zp_i})$ with i runs from 1 to N.  Also from (\ref{kk}) it follows that $f(\cp\zp)\sim {{(\cp\zp)}^n} $.  So when we evaluate the expectation value of WK in this basis, in the expression of W we see that       $\lan\cpm\zpm\ran = 0$ as on a closed contour the angular integral will vanish. in the inner product, the other term $\cp\zm + \zp\cm$ expresses nothing but the hamiltonian of which f is an eigenfunction. 
So if  $ {\psi_n}$  are the set of functions which  diagonalizes K we can write (\ref{partioww}) as
\be
\displaystyle\sum_n {\rm log}\lan{\psi_n}| (1+e^{-2\pi R \b \mu}\hat{W}\hat{ K}) |{\psi_n}\ran ,
\la{prody}
\ee
where
\ber
\hat{W}{K_+}{K_-}f(\cp\zp) 
&=& \int  [d\cm][d\zm]
e^{  i (\cp \zm +\cm\zp)}[d\cp^\prime][d\zp^\prime] e^{ -iq(\cp^\prime \zm +\cm\zp^\prime)} \n
& & (1+{\f{ \cp^\prime \zm +\cm\zp^\prime }{\mu_B^2}} )      
  f({\cp^\prime}{\zp^\prime}).
\la{kkw}
\eer
 As the expression depends on $(\cp \zm +\cm\zp)$ which is the expression for free hamiltonian ${H_o}$ so comparison with (\ref{pa}), Fredholm determinant is expected to be given by  
\be
{\cal{Z}}(\mu) = {\rm det}(1+ e^{-\tpr\b(\mu+H_0 )-\log(1-{\f{2{H_o}}{\mu_B^2}})}).
\la{paw}
\ee
This is,we are going to analyze and evaluate in the next part of this section.
\subsection{Evaluation of the thermal partition function}
In this section we are going to study type 0A MQM in the presence of  D brane with time t compactified on a circle, evaluate and analyze the free energy.  In the absence of the brane  when we compactify string theory on a circle of radius R, in  the dual MQM the  Schrodinger equation have periodic solution i.e ${\psi}(t) ={\psi}(t+2{\pi}R)$, which implies $E={\frac{n}{R}}$.  Now consider the theory with  D brane which can be accomplished by including a macroscopic loop operator localized at $ t={t_o}=0\equiv \tpr$ (say) to the action. From previous discussion it follows that in the   presence of the operator   Schrodinger equation will have well defined solution only in the region $0\leq t \leq \tpr$  when discontinuity occur at the respective point and we have $\psi({\tpr-\epsilon})\ne \psi({\tpr+\epsilon})$  in the limit $\epsilon \rightarrow 0$.  This  is consistent with the fact that the presence of a spacelike brane breaks the winding symmetry and apparently the theory correspond to that of an open string. At the end we will see how the closed string scenario arise in this picture.  Now in a compact time we must have the  condition  ${\psi}(t) ={\psi}(2{\pi}R + t)$.  So effectively we can view the theory as  MQM on a line of length ${\tpr}$ with two ${\delta}\textendash{\rm potential}$ along with the operator $\hat{W}$ (where one is the image of the other, superimposed) placed at its two ends.  When we cross the boundary on either side, situation repeats \footnote{note it never implies periodicity, its just similar to the situation of an open string in 2D with  Dirichlet boundary condition in compact direction and identification of the matter direction with  t.  It winds along the circle m times although ends are not identified.  The open string which wraps m times a circle of length $2\pi{R^\prime}$ with $2\pi m {R^\prime} = \tpr$, we can define same theory on either of the slices $2\pi (n-1) R \le t \le 2\pi n R$, crossing the boundary of the slice  implies going back from that end of the string to the other and hence the situation repeats}, i.e we can define the theory on any of the  slices $2\pi(n-1)R \le t \le 2\pi n R$. So, extending the equation we did in(\ref{schrodingerw}), effectively we have the time dependent Schrodinger equation with double delta potential well, as in the case of a circle:
\begin{eqnarray}
\lbrack \, \, i {\f{\p}{{\p} t}} 
&-& \lbrace {\delta}(t)+{\delta}(t-2{\pi}R)\rbrace W(\hcpm\hzpm(t),{H_o})  \rbrack {\Psi}(\cpm\zpm,t)\nonumber\\
&=& {\mp} i \left\lbrack
 \zpm{\f{\p}{\p \zpm}}+{\cpm}{\f{\p}{\p\cpm}}+1 \right\rbrack {\Psi}(\cpm\zpm,t) ,
\label{double}
\end{eqnarray} 
  We have the discontinuity
\begin{eqnarray}
{\psi}({\epsilon}) - {\psi}({-\epsilon}) 
&=& {\hat{W}}{\psi_o}(t=0) \nonumber\\
{\psi}({2{\pi}R}+{\epsilon}) - {\psi}({2{\pi}R-\epsilon}) 
&=& {\hat{W}}{\psi_o}(t={2{\pi}R}).
\label{discontinuous}
\end{eqnarray}
In order to evaluate the partition function we must need to know what is the right Hilbert space describe the wave function $\psi$ on the circle. This is because we know that the Hilbert space $\lbrace |E\ran\rbrace $ and  $\lbrace |E \pm ni\ran\rbrace $  cannot be mapped to each other. Now, on a circle, the outside region of  the double delta well as described in (\ref{double}) is squeezed to a point, while in the region between the two well, i.e $\epsilon <t<\tpr-\epsilon$, we can think the wave function is the free fermionic wave function, $|E\ran$. So the partition function (\ref{paw}) is given by  the transition amplitude 
\begin{eqnarray}
\lim_{\epsilon\to 0} \,\lan{\psi_o}{(\epsilon)}|\psi_> (\tpr - \epsilon)\ran
&=& \lan{\psi_o}\lbrack {e^{-{\b\int_{0}^{\tpr}} dt{\hat{ H}_o}}}    {e^{{\hat{W}}({t_o}) }}\rbrack {\psi_o}\ran \n     
&=& {{\rm Tr}_{\psi_o^E} } \lbrack {e^{-{\tpr\beta}{\hat{ H}_o}}}   \lbrack {e^{{\hat{W}}({t_o}) }}\rbrack\  \rbrack\nonumber\\
&=& {{\rm Tr}_{\psi_o^E} } \left\lbrack {e^{-{\tpr\beta}\lbrack {\hat{ H}_o}-{\frac{1}{\tpr\beta}}\hat{W}({t_o})    \rbrack}} \right\rbrack,
\label{partition2}
\end{eqnarray}
where ${{\rm Tr}_{\psi_o^E} }$ implies the summation over all free fermion eigenfunctions and $\psi_> $ corresponds to the wave function exactly at $t=0\equiv\tpr$.  As $\b \rightarrow \infty$ at double scaling limit, so inside the partition function we can replace it by $\psi_> \rightarrow e^{-W{(t_o)}}{\psi_o}\,$  \footnote{In order to reach from the 1rst to 2nd step  in (\ref{partition2})we utilize the fact that we can scale the time $t\rightarrow \b t$ so that the term with the macroscopic loop operator $\int dt W(t) \delta(t-{t_o})$  will get a factor ${\f{1}{\b}}$. Hence in the double scaling limit where $\b \rightarrow \infty$  and with Euclidean time, we can lift up the  term to the exponential and the exponent gives an exact expression what we have obtained from the path integral (\ref{partin}) .  Also following the discussion of section 2 we can directly add $ W(t_o)$ in the expression of hamiltonian in Euclidean time to get the expression (\ref{partition2}) }.   Note when we consider Schrodinger equation, the contribution from the term  ${\frac{1}{\tpr\beta}}\hat{W}({t_o})$  in the expression of hamiltonian (\ref {partition2}) at double scaling limit will not be negligible due to a transformation of variable  which leads to the physics at the vicinity of the top of the potential, as we discussed section  $2.1$  and can be found \cite{KIR}.  From (\ref{macr}) we know  that in a single fermionic state the presence of D brane implies implies the insertion of the following operator
\be
{e^{{\hat{W}}({t_o}) }} = 1+ {\f{\hcp\hzp +\hcm\hzm +\hcm\hzp +\hcp\hzm}{\mu_B^2}}.
\la{macroscopicexpression}
\ee
Here first we will evaluate the partition function for single fermionic d.o.f in order to understand the behaviour of the system in the presence of a brane. Next we will derive the grand canonical partition function.
\vskip0.5mm
\textbf{Imposition of the constraint}
Now following the discussion in section 2 we can represent the  wave function  at $t=\tpr-\epsilon$  in (\ref{partition2})  with the operator $\hat{W}$ expressed either in terms of $\hcp,\hzp$ or  $\hcm,\hzm$ representation. This is the implication of constraint(\ref{impose}), (\ref{imposedcondition}).  We have shown in the Appendix that $\lan \cp\zp|\hcp\hzp|\cp\zp\ran$ and $\lan \cm\zm|\hcm\hzm|\cm\zm\ran$ diverge.  So we must express $\hat{W}$ as 
$W({\hcm\hzm,H_o})$ for the basis  $|\cp\zp,E\ran$  basis and vice versa. So applying (\ref{orthomod}) we can write  (\ref{partition2}) as:
 \be
{{\rm Tr}_{\psi_o} }\left \lbrack{e^{-{\tpr\beta}\lbrack {\hat{ H}_o}-{\frac{1}{\tpr\beta}}{\rm log}(1-{\f{2\hat{H}_o}{\mu_B^2}})\rbr} }\right\rbrack
= {{\rm Tr}_{\psi_o} }\left \lbrack e^{-\tpr \beta H_o^\prime } \right\rbrack,
\la{effectivepartition}
\ee 
Where 
\be 
\hat{H}_o^\prime = {\hat{ H}_o}-{\frac{1}{\tpr\beta}}{\rm log}(1-{\f{2\hat{H}_o}{\mu_B^2}}) \quad;\quad {E_o^\prime}={E}-{\frac{1}{\tpr}}{\rm log}(1-{\f{2{E}}{\mu_B^2}}),
\la{heprime}
\ee
with $E_o^\prime$ is the eigenvalue of $H_o^\prime$(note that we omitted $\b$ factor from the expression of ${E_o^\prime}$ following the discussion of section 2, which can be done at double scaling limit by redefinition of the variable).  Before any further analysis let us make the comment that here we have considered the  macroscopic loop operator in the  NS NS sector.  For any other sector we can use analysis of section 2, expressing W in terms of $\hcm.\hzm (\hcp,\hzp)$  in $\cp,\zp(\cm,\zm)$ representation  and using (\ref{orthomod})  to express $W({t_o})$ complete in terms of ${H_o}$ within the trace.  Although we will have a very different expression of ${H_o^\prime}$ but the analysis will remain same. Also if we did not apply this  condition (\ref{impose}), note we will have the term $\lan \hcm\hzm \ran_- (\,\, \lan \hcp\hzp \ran_+ \,\,)$ in the partition function from the expression of $\hat{W}$. This gives rise to an infinite contribution  to the partition function $\lim_{r\to \infty} r{e^{i\phi(E-i)}}$  (where $\phi$ is the phase of the wave function) and so the partition function diverge. This is the signature of the presence of an unphysical degree of freedom leads to  instability of the system due to nonconservation of MQM hamiltonian, which we have removed by our choice of constraint.. 
  Now according to the discussion of section 2, at the double scaling limit the partition function (\ref{effectivepartition}) can be expressed as the sum over ${E_o^\prime}$the eigenvalue of ${H_o^\prime}$ as:
 \be
{{\rm Tr}_{\psi_o} }\left \lbrack e^{-\tpr \beta H_o^\prime } \right\rbrack
=\sum_E \left \lbrack{e^{-{\tpr\beta}\lbrack {E}-{\frac{1}{\tpr}}{\rm log}(1-{\f{2{E}}{\mu_B^2}})\rbr} }\right\rbrack
\la{partitionenergy}
\ee  
 Now note that ${E_o^\prime} $,the eigenvalue of ${ H_o^\prime }$, has branch cut at $E={\frac{\mu_B^2}{2}}$ so we need to subtract a small cut-off ${\rm log} \epsilon$ in order to have an well defined expression of the energy  and after subtraction $E\rightarrow{E^\prime}$ is an one to one mapping,  Also note at the singular point, $ E={\f{1}{2}}{\mu_B^2}$, $e^{-\tpr\b{E^\prime}}$  is trivially zero and so it will not contribute any pole to integrand.  Now the string theory compactified at time interval $2{\pi}R$ is described by the grand canonical partition function of fermion at finite temperature 
${\frac{1}{\tpr}}$ and chemical potential ${\mu}$. So the free energy ${\CF}={\rm log}{\cal{Z}}$ in the presence of Dbrane is given by  
\begin{equation}
{\CF}({\mu})={\int_{\infty}^\infty}  dE {\rho}(E)log(1+{e^{-{\beta}({\mu}+{E^\prime}(E))}}),
\label{partition6}
\end{equation}
where$\rho(E)$ is given in (\ref{density})
\begin{equation}
e^{i\phi_0(E)}
=  R(E)={\f{ \Gamma( iE + 1/2) }{\Gamma( -iE + 1/2)}}.
\label{dense}
\end{equation}
Now we can calculate free energy
${\cal{F}}(\mu, R)= {\rm log}{\cal{Z}}(\mu, R)$ as.
\begin{equation}
{\cal{F}}(\mu, R)=  \int_{-\infty}^\infty dE\rho(E) \log\lbrack1+e^{-\beta(\mu+{E^\prime}(E))}  \rbrack.
\label{freet}
\end{equation}
with the density (\ref{density}),and from (\ref{partition6},\ref{heprime}) the free energy is given by 
\begin{eqnarray}
\CF(\mu, R) 
&=& -{\f{1}{2\pi}}\int d{\phi_0} (E)
{\rm log }\left( 1+{e^{-\tpr\beta({\mu}+ {E^\prime}(E))}}\right)\nonumber\\
&=& - R {\int_{-\infty}^{\infty}} d{E^\prime} {\f {\phi_0 (E({E^\prime})) } { 1+e^{-\tpr\beta (\mu+ {E^\prime}(E)) )}}} \nonumber\\
&=& -i {\sum _{r=n+\hf > 0}
\phi_o(E({E^\prime}= {\f{ir}{\b R}}-\mu))}\nonumber\\
&=& -i {\sum _{r=n+\hf > 0}}
{\phi_o^\prime}( {\f{ir}{ R}}-\mu),
\label{expressiondual1}
\end{eqnarray}
where we have
\be
{\phi_0^\prime} ({E^\prime})={\phi_0} (E),
\la{phirelation}
\ee
Also in order to go from first step to second step in (\ref{expressiondual1}), we followed the same steps as in (\ref{fred}), did a partial integration.
So from the expression of the density of the energy eigenstates (\ref{density}) (ignore the $\Lambda$  factor ) this implies the number of energy eigenstates between E  to  E+dE is same as that of between ${E^\prime}$   to  ${E^\prime}+ d{E^\prime}$. So the partition function on a circle in the presence of D brane corresponds to a deformation in static Fermi sea where the deformation is expressed as  $E=-\mu\,\Rightarrow\,E^\prime =-\mu$ with all the energy eigenstates are in one to one mapping.  Note that the partition function is getting contribution only from the deformation of Fermi surface instead of excitation modes.  Finally the expression of the free energy $\CF(\mu, R)$ in (\ref{expressiondual1}) along with (\ref{phirelation}) suggests the following
\begin{eqnarray}   
{{\rm Tr}_{\psi_o} }\left \lbrack{e^{-{\tpr\beta}\lbrack {\hat{ H}_o}-{\frac{1}{\tpr\beta}}{\hat{W}}({t_o})    \rbrack}}\right \rbrack
&=& {{\rm Tr}_{\psi_o} } \lbrack {e^{-{\tpr\beta}\lbrace{H_o}-{\frac{1}{\tpr\beta}} {\log}(1-{\frac{2{H_o}}{\mu_B^2}} )  \rbrace     }}     \rbrack \nonumber\\
&=& {{\rm Tr}_{\psi_o} } \lbrack {e^{-{\tpr\beta}\lbrace{H_o}-{\frac{1}{\tpr\beta}}f({H_o} )  \rbrace     }}     \rbrack \nonumber\\
&=& {{\rm Tr}_{\psi_o} } \lbrack {e^{-{\tpr\beta}({H_o^\prime} )   }}     \rbrack\nonumber\\
&=& {{\rm Tr}_{\psi_o^\prime} } \lbrack {e^{-{\tpr\beta}({H_o})    }}     \rbrack,
\label{partition3}
\end{eqnarray}
where 

\begin{equation}
f({H_o})={\log}(1-{\frac{2{H_o}}{\mu_B^2}} ) {\quad;\quad}{H_o^\prime} = {H_o}-{\frac{1}{\tpr\beta}}f({H_o}).
\label{def2}
\end{equation}

and in the last step we made a transformation from the basis ${\psi^\pm_o}(E)={e^{\mp{i\phi_o}(E)}}{e^{-iEt}}{{(\cpm\zpm)}^{\pm iE-{\frac{1}{2}}} }\rightarrow
{\psi^\pm_o}({E^\prime}) ={e^{\mp{i\phi_o^\prime}({E^\prime})}}{e^{-i{E^\prime}t}}{{(\cpm\zpm)}^{\pm i{E^\prime}-{\frac{1}{2}}} }$ with
 ${E^\prime}=  E - {\frac{1}{\tpr}}{\log}(1-{\frac{2E}{\mu_B^2}} )$ and also ${\phi_o^\prime}({E^\prime})= {\phi_o}(E)$. Note as we discussed in section 2, although the contribution from the macroscopic loop operator has a factor ${\f{1}{\b}}$ however in the double scaled hamiltonian it cannot be ignored and the shifted energy will be given by $E^\prime$.  Above expression implies that the Type 0A MQM on a circle in the presence of a D brane  can be viewed as a free theory with the free hamiltonian ${H_o}$ with the wave function replaced by the above one.  This point will be relevant in section 5.
Note the relation (\ref{phifrerelation},\ref{genusexpansion}) can be expressed as
\be
2{\rm sin }{\frac{\p_\mu }{2\b R} }\cdot\CF(\mu ) = \phi_o ^\prime(-\mu) .
\la{phibranerelation}
\ee
Also  note that from nonlinear relation between the effective hamiltonian ${H_o^\prime}$ and the free hamiltonian ${H_o}$ its evident that in the genus expansion of free energy in the relation (\ref{genusexpansion}) will have both odd and even powers of $\mu$( this is because the new free energy corresponds to replacing $\mu \rightarrow E(E^\prime){|_ {E^\prime = \mu}}$ which follows from (\ref{partition6})).  This is the signature of the presence of surface with boundary which is the implication from MQM/string theory duality. 

\section{Fermionic scattering and semiclassical analysis}
\setcounter{equation}{0}

In this section we will study scattering of fermions in the presence of D brane and tachyonic background at quasiclassical limit.
First we briefly review the formal theory.

\subsection{Fermionic scattering: the review} 

The scattering amplitude is given by
\be 
S= \lan \b ,{t\rightarrow \infty }|\alpha,{t \rightarrow -\infty }\ran,
\la{S}
\ee
where ${\alpha}$ and ${\beta}$ denotes the incoming and outgoing state.  As the single incoming and outgoing  state is given by  $|\cp\zp\ran$ and $|\cm\zm\ran$ respectively \cite{lectjevicki}, so
\be
S=\lan \cm\zm,{\rm out}|\cp\zp,{\rm in} \ran.
\la{scat}
\ee
Now note that  $\cp\zp$ and $\cm\zm$ representations are related by a unitary operator $\hat S$, which in our case is nothing but the Fourier transformation 
on the complex plane.  Recall, the energy eigenstates in absence of the D brane  are given by (\ref{freesolution}).  The wave functions in $(z_+,\bar z_+)$ and
$(z_-,\bar z_-)$ representations are related by 
\ber
\psi_-(z_-,\bar z_-) 
&=& \hat{ S }{\psi_+}(z_-,\bar z_-) \n
&=& \int dz_+ d\bar z_+ K(\bar z_-,z_+) K(z_-,\bar z_+)\psi_+(z_+,\bar z_+),
\la{scatter}
\eer
where $K(z_-, z_+) = {1\over \sqrt{2\pi}}e^{iz_- z_+}$. Acting on energy eigenstates, we have
\be
 \hat{S} \psi_+^E = {\cal {R}}(E) \psi_-^E,~~~~ 
 {\cal {R}}(E) = {\f{\Gamma(iE+{\f{1}{2}})}{ \Gamma(-iE+{\f{1}{2}})}}  .
\la{s}
\ee
The factor $\CR(E)$ is a pure phase  
 \be
\overline{\CR(E) } \CR(E) = \CR(-E) \CR(E)=1 ,
\la{r}
\ee
which proves the unitarity of the operator $\hat{S}$.  Now in absence of the D brane,  wave function (\ref{freesolution}) evolve according to  free hamiltonian ${H_o}$, so from orthonormality of the wave functions (\ref{scat}) is given by $\CR(E){\delta}({E_+}-{E_-})$. The operator $\hat S$ relates the incoming and the outgoing waves and therefore can be interpreted as the fermionic scattering matrix.  The factor $\CR(E)$ is identical to the the reflection coefficient. This condition can also be expressed as the orthonormality of in and
out eigenfunctions
\be
\langle  \Pim{E_{{-}}}|K|  \Pip{E_{{+}}}\rangle =\delta(E_+-E_-),
\la{inner}
\ee
with respect to the scalar product.
We usually absorb the factor$\CR(E)$ in phase by defining 
\be
e^{i{\phi(E)}}= \CR(E)
\la{phaserelation}
\ee 
to make the wave function biorthogonal, where ${\phi(E)}$ is the phase of the incoming and the outgoing wave function(\ref{freesolution}).

\subsection{Fermionic scattering in the presence of D brane: The imposition of the constraint} 

 Here we consider, the theory of scattering, in  the presence of D brane. For a single fermionic state, (\ref{scat}) is given by
\be
\lan \cp\zp,t = \infty|{e^{\hat{W}(t)}}| \cm\zm,t = \infty \ran = \lan \cp\zp,{\rm out}|(1-{\f{\cp\zp+\cm\zm - 2{H_o}  }{ \mu_B^2} }  )| \cm\zm,{\rm in}\ran.
\la{wsc}
\ee
Now according to (\ref{imposedcondition}) W will be expressed either in $\cp\zp$ or $\cm\zm$ mode
\ber
\lan {z_+},\ep,{\rm out}|e^{{\hat{W}}({t_o})}| {\zm},{E_-},{\rm in} \ran
&=&\lan {\zp},\ep|(1+{\f{2\cm\zm - 2{H_o}  }{ \mu_B^2} } { )_{t={t_o}}}| {z_-},{E_-}{\rm in} \ran\n
&=&\lan {z_+},\ep|(1+{\f{- 2{H_o}  }{ \mu_B^2} }  )| {z_-},{E_-}\ran\n
&=&\lan {z_+},\ep|(1-{\f{2E  }{ \mu_B^2} }  )| {z_-},{E_-}\ran\n
&=& \CR(\ep){e^{log(1-{\f{2E  }{ \mu_B^2} }  ) }}\d(\ep -{E_-}),
\la{wscat}
\eer
where in the 2nd step we have used $\lan {\cp\zp},\ep ; {\rm out}|\hcm\hzm | {\cm\zm},{E_-},{\rm in} \ran = 0 $  from (\ref{orthogon}) .
Now  the presence of the D brane will modify the phase of the outgoing state over the incoming, which is given by the factor ${e^{log(1-{\f{2E  }{ \mu_B^2} }  ) }}$. So for the change of phase $ \delta\phi(E)$ we can write
\be
e^{-i\f{\delta \phi(E)}{2}}={e^{log(1-{\f{2E  }{ \mu_B^2} }  ) }}
\la{changephase}
\ee
Now in (\ref{changephase}) using the relation (\ref{phaserelation}) will leave us with the amplitude $e^{-i\f{\delta \phi(E)}{2}}$.  In \cite{mnzz} its explained that the complex phase in the wave function is the signature of tunneling and we can presume the above factor accounts for the same. Also note that instead of the above macroscopic loop operator in  NS  sector, if we took the  macroscopic loop operator in some other sector given by 
 ${W^\prime}({\f{\cp\zp,\cm\zm,{H_o}}{\mu_B^2}})$  according to the discussion in Appendix we will have the phase shift of the outgoing state ${\f {\phi(E)}{2}} + i{W^\prime (1-{\f{ 2E   }{ \mu_B^2} }  )}$. 
 \vskip 0.5mm
 Lets  consider scattering of a tachyonic state (which are being created from the action of ${(\hcpm\hzpm)}^n$ on fermionic ground state) from D brane.  This is better understood from the collective field theory where scattering to a single tachyonic state with energy E is given by $\sim U(E)$ where U(E) is given by (\ref{up}). This supports the fact that the D brane act as a coherent source of closed strings.
 Now we consider the classical limit $\b \rightarrow \infty$.    
At this limit the ground state of MQM is obtained by filling all energy levels up to some fixed Fermi energy which we choose to be $E_F=-\mu$.  Quasiclassically every energy level corresponds to a certain
trajectory in the phase space of $\cp\zp,\cm\zm$ variables and they are separated by a factor ${\f{1}{\b}}$.  The Fermi sea can be viewed as a stack of all classical trajectories with $E\le E_F$ and the ground state is completely
characterized by the curve representing the trajectory of the 
fermion with highest energy $E_F$. For the Hamiltonian  ${H_o}$
all trajectories are hyperboles $\cp\zm  + \cm\zp = -E$ 
and the profile of the  Fermi sea is given by 
\be
\cp\zm  + \cm\zp = -\mu.
\la{fermi}
\ee
First consider the theory without D brane. Then the low lying collective excitations are
represented by deformations of the Fermi surface, 
\be
\cp\zm + \cm\zp  = M(\cp\zm + \cm\zp ).
\la{ferm}
\ee
In order to study the scattering with such deformed background we will follow the analysis of \cite{timedependentbgd}.  The perturbed wave functions are related to the old ones by a phase factor
\be
{\psi^E_{\pm}}(\cpm\zpm)=e^{\mp i {\vp_\pm} (\cpm\zpm;E)}{\psi^E_{\pm}}(\cpm\zpm),
\la{phas}
\ee
whose asymptotics at large $\cpm\zpm$ characterizes the incoming/outgoing
tachyon state. We split the phase into three terms
\be
\vp_\pm  (\cpm\zpm;E) 
= V_\pm(\cpm\zpm) +{\f{1}{2}} \phi(E) + v_\pm(\cpm\zpm;E),
\la{wave}
\ee
where the potentials $V_\pm$ are fixed smooth functions vanishing at
$\cpm\zpm=0$, while the term $v_\pm$ vanishing at infinity and the
constant $\phi$ are to be determined.  Now in order to understand the time-dependent profile of Fermi sea  first consider the situation in the absence of the brane as described in \cite{timedependentbgd}.
\be
\langle \Pim{E_-}|\Pip{E_+}\rangle=\CN {e^{-i\phi}}
\int_{0}^\infty {\f{d\cp d\cm d\zm d\zp } {\sqrt{\cp\zp} \sqrt{\cm\zm} } }{e^{i(\cp\zm+\zp\cm)}}
e^{- i {\vp_+}(\zp)-i{\vp_-}(\zm) } 
{(\cp\zp)}^{iE_-}{(\cm\zm)}^{iE_+},
\la{scatt}
\ee
where $\CN$ is the normalization.
At $\beta  \rightarrow \infty $ Fermi profile can be obtained from saddle point approximation which is given by
\be
\cp\zm +\zp\cm =-E_\pm+(\zpm \p _\pm + \cpm{\overline{\p} _\pm })\vp_\pm(\cpm\zpm).
\la{profileo}
\ee
So following \cite{timedependentbgd} it appears the perturbed state will be an eigenstate of the deformed  hamiltonian $H = {H_o}+{H_p}$ where $H_p$ is given by
\be
{H_p}= (\zpm \p _\pm + \cpm{\overline{\p} _\pm })\vp_\pm(\cpm\zpm; H)
\la{hp}
\ee
Now in the presence of the D brane scattering matrix element will be given by 
$$\lan \cp\zp,\ep,{\rm out}|e^{\hat{W} ({t_o})}| ,{E_-},\cm\zm, {\rm in}\ran$$. 
S\textendash matrix element is expressed as
\ber
S_{perturb}
&=    &  {e^{-i\phi}
\CN\int\limits_{0}^\infty { \f{  d\cp d\cm d\zm d\zp }{ \sqrt{\cp\zp} \sqrt{\cm\zm} }}{e^{i(\cp\zm+\zp\cm)}}
 e^{-i{\varphi}{(\cm\zm)} }{(\cm\zm)}^{iE_{-}} }\n
&     & \lbr e^{-i{\int_{t_o}^\infty}{\hat{H}_o}}  \lbr e^{-i{W_{\rm proj}}({t_o})} \rbr e^{i{\int_{-\infty}^{t_o}}{\hat{H_o}}}\rbr 
 e^{-i{\varphi_+}(\cp\zp) }{(\cp\zp)}^{iE_{+}}\n
&\sim & e^{-i\phi}
\CN \int\limits_{0}^\infty{\f {d\cp d\cm d\zm d\zp } {{\sqrt{\cp\zp} }\sqrt{\cm\zm} }}{e^{i(\cp\zm(t)+\zp\cm(t))}}\n
&     & e^{-i{\varphi}(\cm\zm(t)) }{(\cm\zm(t))}^{iE_-} 
\lbr  e^{iW(t)(\cpm\zpm,{H_o})}\rbr
 e^{-i{\varphi_+} (\cp\zp(t)) } {(\cp\zp(t))}^{iE_+}\n,
\la{pertu}
\eer
where by ${W_{\rm proj}}$ we imply the macroscopic loop operator expressed in either $\cp\zp$ or $\cm\zm$ mode, as we have been doing using the constraint(\ref{imposedcondition}).
So in the presence of the D brane, in the classical regime  from (\ref{pertu}) we can write the Fermi profile in the presence of D brane as
\be
\cp\zm +\zp\cm =-E_\pm+(\zpm \p _\pm + \cpm{\overline{\p }_\pm }) \vp_\pm(\cpm\zpm) + (\zpm \p _\pm + \cpm{\overline{\p }_\pm })W(\cpm\zpm,E)
\la{fermiprofile}
\ee
The perturbed hamiltonian for the deformed state is given by
\be
{H_p^\prime}= {H_p}+(\zpm \p _\pm + \cpm{\overline{\p} _\pm })W(\cpm\zpm; H)
\la{hpp}
\ee
So  we see in the presence of D brane Fermi profile develops instability.
\section{Perturbation by momentum modes}
\setcounter{equation}{0}


In this section we will consider type 0A matrix model with the time t compactified on a circle of radius R, perturbed by momentum modes ${V_{\frac{n}{R}}}$ in the presence of  D brane and evaluate the partition function. Before going to the theory with D brane, in the next two subsections we will briefly review, the wave function in a background perturbed by momentum modes and Lax formalism

\subsection{Review of the theory with momentum mode perturbation}

 Here first we briefly review the scenario without  D brane and then study what happens when we consider the theory in the presence of  D brane. The tachyon modes of the closed string theory are the asymptotic states of collective field theory \cite{stringfield}. The discrete tachyonic  operator $ \CT_{n} \sim  \int _{\rm world  sheet } e^{\pm inx/R}  e^{(|n|/R-2)\phi}$ corresponds to the following  operator in matrix model\cite{lectjevicki,classicallimit},
\begin{equation} 
{V_{{\pm}n/R}}= e^{-{\f{n}{R}}t}{{(\cpm\zpm)}^{n/R}}.
 \label{per}
\end{equation}
These operators creates a discrete tachyonic state of momenta ${\f{n}{R}}$ over the matrix model ground state and are  periodic in Euclidean time.
\be
\lbr {H_o},{V_{{\pm}n/R}}\rbr = \mp i{\f{n}{R}}{V_{{\pm}n/R}}\quad;\quad k\geq 1.
\la{eigenfunction}
\ee
So ${V_{{\pm}n/R}}$ shift the energy  $E\rightarrow E\mp i{\f{n}{R}}$ cause a time-dependent perturbation to Fermi sea.  The perturbed state in general can be expressed as 
\be 
 \Psi_{\pm}^E = e^{\mp i \varphi(z_\pm \bar z_\pm;E)} \psi_{o\pm}^E \equiv {\cal W}_\pm \psi_{o\pm}^E ,
\la{dreswave}
\ee
 where the phases $\varphi_\pm$ have Laurent expansion
\be
{\varphi_\pm}(\zpm\overline{z}_\pm;E) = {\f{1}{2}}\phi(E) + R\sum_{k\geq 1} t_{\pm k} {(z_\pm\overline{z}_\pm)}^{k/R} -R \sum_{k\geq
1} {\f{1 }{k}} v_{\pm k} ({z_\pm} \overline{z}_\pm)^{-k/R}.
\la{varphi}.
\ee
$t_{\pm k}$ parametrize the asymptotic perturbation by momentum modes of
NS-NS scalars, corresponding to the operator introduced  (\ref{per}),  Note the above wave function  asymptotically behave as 
 \be
\Psi^E_\pm (\cpm\zpm)
\sim {(\cpm\zpm)} ^{\pm iE-\hf}\  e^{ \mp \hf i \phi(E) }\
e^{i U_\pm (\cpm\zpm)} \quad ;\quad U_\pm (\cpm\zpm) = \sum_{k\geq 1} {|\cpm\zpm|}^{\f{k}{R}}.
\la{asymptotic}
\ee
From  the above its evident that tachyonic  perturbation can be achieved by  deforming the integration measures  $d\lbr\cpm\zpm\rbr$ to \cite{intflow}
\be
[d\cpm\zpm] \rightarrow [d\cpm\zpm]  {\rm exp}\left(\pm i{U_\pm}(\cpm\zpm)\right). 
\la{mes}
\ee
Extending the discussions of section 3, these wave functions  (\ref{dreswave}) diagonalizes the deformed kernel (\ref{mes}).  While the perturbed wave function evolves in time with ${H_o}$, but it can be seen as the stationary state w.r.t an effective hamiltonian $H= {H_o} + {H_p}(H)$, where the expressions for perturbed hamiltonian ${H_ p }$ from semiclassical analysis is obtained in \cite{timedependentbgd} .  The partition function is given by ${\rm Tr}e^{-2 {\pi}R \b H}$, following section 3 which can also be expressed  as Fredholm determinant.  We have the free energy$\CF = -i \sum _{r\ge 1/2} \phi \left( i r/R-\mu\right)$  where $\phi(E)$ is the phase described by (\ref{varphi}).  It satisfies the equation 
\be
\phi (-\mu)
= 2  {\rm sin } \left({\f {\p_\mu} {2 \b R}}\right)\CF(\mu, R) ..
\la{phase}
\ee

\subsection{Review of Lax Formalism}
Here we will briefly review the Lax formalism in the context of Type 0A matrix model. Consider the operator $(\hat z_\pm \hat{\bar z}_\pm)$  which can be represented as shift operators $\hat\omega^{\pm 1}$, where $\hat\omega$ acts on energy
eigenstates as $\hat \omega^{\pm1}\psi_\pm^E = \psi_\pm^{E\mp i}$.  We have $\hat\omega=e^{-i\partial_E}$ shifts the variable E by $i$.
The operators $\oo$ and $\hat{E}$ satisfy the Heisenberg-Weyl commutation relation
\be
 [\oo,-\hat{E}]= i\oo, \qquad
[\oo^{-1},-\hat{E}]= - i\oo^{-1}.
\la{comcom}
\ee
Now let us consider the representation of these commutation relations in the perturbed theory.  The  dressing operators $\CW_\pm$ 
(\ref{asymptotic})are now   exponents of 
series in $\oo$ with $\hat{E}$-dependent coefficient
\be
\hat \CW_\pm = e^{ iR
\sum_{n\ge 1} t_{\pm n} \oo^{n/R}}\ 
e^{\mp i\phi(E)}
e^{ iR\sum_{n\ge 1} v_{\pm n}({E})\  \oo^{-n/R}}.
\la{w}
\ee
The operators
\ber
L_+ 
&=& \CW_+ \oo 
   \CW^{-1}_+, \quad L_- = \CW_- \oo ^{-1},
   \CW^{-1}_-,\n
M_+ 
&=& -{\CW_+ }\hat{E}
   {\CW^{-1}_+} \quad {M_-} = -{\CW_-} {\hat{E}},
   \CW^{-1}_-.
\la{lm}
\eer
known as Lax and Orlov-Schulman operators satisfy the same commutation  relations  as the operators $\oo$ and $\hat{E}$
\be
[L_+, M_+] =  i L_+ \quad , \quad [L_-, M_-] = - i L_- .
\la{laxcommutator}
\ee
The Lax operators $L_\pm$  represent the canonical coordinates $\hcpm\hzpm$ in the basis of perturbed wave functions
\be
\langle E |e^{\pm i{\f{\phi_0}{2}}}\hat \CW_\pm  
 L_\pm |\cpm\zpm\rangle =\langle E |e^{\pm i{\f{\phi_0}{2}}}\hat {\CW}_\pm  \hcpm\hzpm
 |\cpm\zpm\rangle,
\la{m}
\ee
while  the Orlov-Shulman operators $M_\pm$ represent hamiltonian $H_0 = -
{\frac{1}{2}}( \hcp\hzm+\hcm\hat\hzp)$. 
Therefore the L and M  operators 
 are related also by
 \be
 M_+=M_-,~~~ [L_+,L_-]= 2iM_\pm,~~~\{L_+,L_-\}=2M_\pm^2-{1\over2}.
\la{mla}
\ee
The last identity  is  not  satisfied automatically in the Toda system and represent an additional constraint  analogous to the string equations. The operators $M_\pm$ can be 
expanded as infinite series of the 
$L$-operators. Indeed, as they act to the
dressed wave functions as 
\ber
\langle E |
& & e^{\pm i\phi_0} \hat \CW_\pm \; 
 {M_\pm }|\cpm\zpm\rangle
= \pm i (\zpm\p_{\zpm} + \cpm\p_{\cpm}+ 1)
\Psi^E_\pm (\cpm\zpm)\n
 &=& \left(\sum _{k\ge 1} 
 k t_{\pm k}  {(\cpm\zpm)}^{ k/R}  +\mu+
\sum_{k\ge 1} v_{\pm k} 
 {\cpm \zpm}^{-k/R}\right)\Psi^E_\pm ((\cpm\zpm) ).
\la{mexpression}
\eer
we can write
\be
M_\pm =
   \sum _{k\ge 1}  k t_{\pm k}   L_\pm^{ k/R}+\mu  +
\sum_{k\ge 1} v_{\pm k}  L_\pm^{-k/R}.
\la{mpm}
\ee
In order to exploit the Lax equations 
and the string equations 
we need the explicit form of the two  operators.
It follows from  
that $L_\pm$
can be represented as series of the form
\ber
{L_+ }
&=&  e^{- i\phi/2}
\left(\omega   +\sum_{k\ge 1} a_{ k}
   \omega^{ 1- n/R} \right)e^{ i \phi/2} ,\n
{L_-}
 &=&  e^{ i\phi/2}
\left( \omega^{-1}  +\sum_{k\ge 1} a_{- k} 
\omega^{- 1+ n/R}
 \right)e^{- i \phi/2}.
\la{lplm}
\eer
Recall that the dressing operators ${\cal W}_\pm$ in terms of
$\hat E$ and $\hat \omega$ are of the form
\be
{\cal
{W}}_\pm = e^{\mp i\phi/2} \left( 1+\sum_{k\geq 1} w_{\pm k}\hat{\omega}^{\mp k/R} \right) e^{\mp iR \sum_{k\geq 1} t_{\pm k}\hat {\omega}^{\pm k/R}} 
\la{wrepresentation}
\ee
Studying the evolution laws of the Orlov--Shulman operators, one can find
that \cite{thesisalexandrov}
\be
{\p v_k\over \p t_l}={\p v_l\over \p t_k}.
\la{vttv}
\ee
It means that there exists a generating function $\tau_s[t]$
of all coefficients $v_{\pm k}$
\be
v_k(s)={({\frac{1}{\beta}})^2}\, {\p \log \tau_s[t]\over \p t_k}.
\la{tauvk}
\ee
It is called {\it $\tau$-function} of Toda hierarchy. It also allows
to reproduce the zero mode $\phi$ and, consequently, the first coefficient in the expansion of the Lax operators
\be
e^{\b \phi(s)}={\tau_{s}\over \tau_{s+{\f{1}{\b}}}},
\qquad
r^2(s-{\f{1}{\b}})={\tau_{s+{\f{1}{\b}}}\tau_{s-{\f{1}{\b}}}\over\tau_s^2}.
\la{taur}
\ee
We are going to show that the partition function coincides with $\tau$\textendash function (\ref{tauvk}).  Finally note as the partition function is described in terms of the Fermi level ${\mu}$.   So in the description of Lax formalism we will replace E by ${\mu}$. Now let us discuss  about the integrable flow.
Let us identify the integrable flows associated with the coupling constants $t_n$.
\be 
\p_{t_n} L_\pm  = [ H_n, L_\pm], 
\la{flow}
\ee
where from (\ref{lm}), the operators $H_n$ are related to the
 dressing operators as

\be
H_{n} = ({\p_{t_ n}}\CW_+) \CW_+^{-1} =
 ({\p_{t_n}}\CW_-){ \CW_-^{-1}}.
\la{h}
\ee
it is clear that $H_n = W_+
\oo^{n/R} W_+^{-1} + $ negative powers of $\oo^{1/R}$, which implies  expression of ${H_n}$ can be given by \cite{intflow}
\be
 H_{ \pm n} =  (L_\pm^{n/R}   ) _{^{>}_{<} }
 +{\frac{1}{2}}(L_\pm^{n/R}   ) _{0} , 
\qquad n>0,
\la{hn}
\ee
\be      
 \p_{t_m} H_n -\p_{t_n}H_m - [H_m, H_n]=0.
\la{p}
\ee
 Equations (\ref{p},\ref{h},\ref{hn}) imply that
the perturbed theory possesses the Toda lattice
integrable structure.  The Toda structure implies  an 
infinite hierarchy  of PDE's for the
coefficients $v_n$ of the dressing operators, 
the first of which is
the Toda equation for the phase $\phi(\mu)\equiv \phi(E=-\mu)$
         
\be
i{\f{\p } {\p t_1}}
{\f{\p }{\p t_{-1}} }\phi(\mu)
= e^{i\phi(\mu)-i\phi(\mu-i/R)}
 - e^{i\phi(\mu+i/R)-i\phi(\mu)}.
\la{phiflow}
\ee
\subsection{String theory on a circle with the D brane  in the presence of tachyonic background}

In this section  we are going to evaluate the free energy of type 0A MQM in the presence of D brane and with tachyonic background, in the grand canonical ensemble and try to understand the relevant string theory. Recall in section 3 we have seen that in the absence of the momentum modes, within the time circle $0 \le t \le 2\pi R$ the solution of the  Schrodinger equation corresponds to the free fermionic wave function.  This in the string theory side giving a picture that we have free closed string states along the circle and the coherent states are strongly localized at  $t={t_o}\equiv i{X^o}$ . So with the same view in the presence of tachyonic background, within the circle $0\le t \le \tpr$ the wave function must be  given by (\ref{dreswave}).  

The perturbed wave function, while time dependent w.r.t the free hamiltonian ${H_o}$, it is stationary w.r.t an effective hamiltonian H, similar as discussed in section 5.1, originally obtained in \cite{timedependentbgd}. So lets consider the perturbed MQM  with the effective hamiltonian $H={H_o}+{H_p}(H)$ in the presence of D brane.
First consider the partition function (\ref{parti})  where now we replace  the integration kernels  with the deformed measures (\ref{mes}).  So as a generalization of (\ref{parti}), in the perturbed background,
 the Matrix model partition function in the presence of D brane (\ref{partin}, \ref{partim},\ref{partima} ) will be with the deformed kernel as
\ber
{\cal{Z}}_N(t)
&=&\int\limits _{-\infty}^\infty
 \prod_{k=1}^N
 [d \zp_k ][d \zm_k] [d \cp_k ][d \cm_k] [d{t_k}]{e^{i {t_{n\pm}}{(\cpm\zpm)}^{\nR} }}
 {\rm det}_{jk}\left( e^{i {t_{jk}}\cp_j \zm_{k}}\right)\n 
& &{\rm det}_{jk} \left( e^{-iq \zp_{j} \cm_k}\right)
{\rm det}_{jk}\left( e ^{i{t^{-1}_{jk}} \cp_j \zm_{k}}\right)\ 
{\rm det}_{jk} \left( e^{-iq \zp_{j} \cm_k}\right)\n
& & {\rm exp}\lbr \displaystyle\sum_i {\rm log }(1+{\frac{ \cp_i\zp_i+ \cm_i\zm_i+ \cp_i\zm_i+ \cm_i\zp_i }{\mu_B^2}})\rbr.
\la{parandeform}
\eer
The partition function will be given by Fredholm determinant ${\rm Det}(1+ e^{-\b \mu}WK) )$ (\ref{partioww}) where in order to evaluate the determinant we need to choose the basis which diagonalizes K (i.e (\ref{kk}) with deformed measure as given in (\ref{parandeform})\,)
and we evaluate the expectation value of ${\hat{W}}$ in the same.  In order to evaluate free energy in the presence of D brane we will proceed in the following way. First consider the scenario without D brane.  Recall the expression of free energy  which is expressed in terms of the phase of wave function \cite{intflow,yin}(which is of the same form of (\ref{expressiondual1}), expressed in the absence of brane).  In a perturbed background the phase $\phi (E) $ will be replaced by that of the perturbed wave function (\ref{dreswave}) in the expression of free energy\cite{intflow}. So for the effective hamiltonian H (  $H={H_o}+{H_p}(H)$ where   $H_p$in the semiclassical limit obtained in  (\ref{hp})\, ) of which (\ref{dreswave}) is an eigenfunction, the analysis of section 3.3 implies that free energy of the perturbed system in grand canonical ensemble is given by $\CF = {\rm log}{\cal{Z}}$ with   ${\cal{Z}}= \displaystyle\sum_{N=0}^{\infty} e^{-\tpr\b \mu N}\lb{\rm Tr}e^{-\tpr\b H}{\rb_N}= {\rm Det}\,(1+ e^{-\tpr\b (\mu+H)})$. This is supported from the view of \cite{timedependentbgd} where in the semiclassical regime the explicit expression of ${\cal{Z}}$ is obtained in this form.
The Fredholm determinant (\ref{kk}) in a perturbed background is given by   ${\cal{Z}}$ \cite{intflow}.  In the presence of D brane we have the Fredholm determinant (\ref{partioww}) which in perturbed background is expressed in (\ref{parandeform}).  So as in section 3.3 free energy must be obtained from the thermal partition function in the presence of D brane i.e by insertion of the operator $e^{W(t_o)}$ in the partition function and evaluating the expectation value.  So here in hamiltonian formalism we will evaluate the  grand canonical partition function ${\rm Det}(1+ e^{-\b(H+\mu)})$ in the basis (\ref{dreswave}) with the insertion of the operator and it must be same as the Fredholm determinant (\ref{parandeform}).  Here we are going to show that if we consider the projected theory as described in section 2(i.e the macroscopic loop operator is expressed with either $\cp\zp$ or $\cm\zm$ following (\ref{imposedcondition})), the above mentioned grand canonical partition function  have the integrable structure of tau function of Toda hierarchy. Now the partition function with the momentum modes in the presence of the D brane  is given by the transition amplitude from the initial state ${\cal{W}}{\psi_o}$ to the final state${\cal{W}}^\prime \psi_{o  >}$, where they represent the fermionic wave function before and after being scattered from the D brane and the corresponding  dressing operator is ${\cal{W}}^\prime$ .Now note  that in a compact dimension just before being scattered,  the wave function at $t=\tpr-\epsilon$ must be given by the one at $t=\epsilon$ with a time evolution $\tpr$.  This leads to the identity
\be
{\cal{W}}^\prime \psi_{o>}(t_o) =  {\cal{W}}^\prime(1-W(\hcpm\hzpm,{H_o})){\psi_o}(t_o )=\left(1- W(\hcpm\hzpm,{H_o}) \right){\cal{W}}{\psi_o}(t_o),
\la{coincidence}
\ee
where ${t_o}=0\equiv\tpr$.   Hence the partition function on the circle corresponds to the transition amplitude
\ber
{\cal{Z}}
&=& \lim_{\epsilon\to 0} \,\lan{{\cal{W}}\psi_o}{(\epsilon)}|{\cal{W}}^\prime \psi_{o>} (\tpr - \epsilon)\ran\n
&=& \lim_{\epsilon\to 0} \,\lan{{\cal{W}}\psi_o}{(\epsilon)} | (1-W(\hcpm\hzpm,{H_o}))|{\cal{W}} \psi_{o} (\tpr - \epsilon)\ran\n
&=&{\rm Tr}_ {{\cal{W}}\psi_o} \lb e^{{-\beta}\lbr{\int_{\epsilon}^{2{\pi}R -{\epsilon}} }dt H   + {\int_{-\epsilon}^{\epsilon}} dt  H \rbr+ 
 {\int_{-\epsilon}^{\epsilon}} dt W{\delta}(t)\rb} \rb \n
&=& {\rm Tr}_ {{\cal{W}}\psi_o} \lb e^{{-\beta}\lbr{\int_{\epsilon}^{2{\pi}R -{\epsilon}} }dt   H\rbr} e^ { W(t=0)}  \rb\n
&=& {\rm Tr}_ {{\cal{W}}\psi_o} \lb e^{{-2{\pi}R\beta} H} e^ { W(t=0)}  \rb .
\la{part}
\eer
Where the partition function is evaluated in Euclidean time   and ${\rm Tr}_ {{\cal{W}}\psi_o}$ denotes the trace taken w.r.t (\ref{dreswave})\footnote{In order to reach from the 2nd to 3rd step  in (\ref{part})we utilize the fact that we can scale the time $t\rightarrow \b t$ so that the term with the macroscopic loop operator $\int dt W(t) \delta(t-{t_o})$  will get a factor ${\f{1}{\b}}$ so that in the double scaling limit where $\b \rightarrow \infty$  and with Euclidean time, we can lift up the  term to the exponential and the exponent gives an exact expression what we have obtained from the path integral (\ref{formal})}.  Grand canonical Partition function,${\rm Det}(1+ e^{-\b(H+\mu)}e^{-W(t_0)})$, will be given by the following expression where we will have the contribution from singlet states only
\be
 {\displaystyle\prod_E}\lb1+e^{{-2\beta}{\pi R}(\mu+E)}\lan{\psi_p^E}| e^ { \hat{W}(t=0)} |{\psi_p^E}\ran \rb\n
 ={\displaystyle\prod_E}\lb1+e^{{-2\beta}{\pi R}(\mu+E )}\lan{\psi_p^E}| e^ { \hat{W}} |{\psi_p^E}\ran\rb,
 \la{pari}
\ee
where $\psi_p = \CW \psi_0$.  Note, we could write the above expression for grand canonical partition function only because $\hat{W}$ can be expressed as the direct product of the operators for the single fermionic states. Now following (\ref{macr}) the partition function  (\ref{part}) can be expressed as
\ber
{\displaystyle\prod_E}
& & \lbr 1+e^{{-2\beta}{\pi R}(\mu+E) }\lan{\psi_p^E}| e^ { {\rm log} (1+{\frac{ 2\hcpm\hzpm -2{H_o}}{\mu_B^2}})    } |{\psi_p^E}\ran\rbr\n
&=& {\displaystyle\prod_E}\lbr 1+e^{{-2\beta}{\pi R}(\mu+E) }\lan{\psi_p^E}|   (1+{\frac{ \hcpm\hzpm-2{H_o} }{\mu_B^2}})|{\psi_p^E}\ran\rbr.
\la{pari}
\eer
Now  we have shown in the appendix that $\lan \cp\zp|\hcp\hzp|\cp\zp\ran$ and $\lan \cm\zm|\hcm\hzm|\cm\zm\ran$ diverge.  So we must express $\hat{W}$ as 
$W({\hcm\hzm,H_o})$ for the basis  $|\cp\zp,E\ran$   and vice versa.  Now note according to the commutation relation (\ref{com}) and from the form of the wave function (\ref{dreswave})  $$ \hcm\hzm = {e^{-i{\f{\varphi{(\hat{E+i})}}{2}} }}{\pcp\pzp} {e^{i{\f{\varphi{(\hat{E})}}{2}}  }}$$.
So according to the analysis of Appendix, $\lan{\psi_p^E}| {\frac{ 2 \hcm\hzm}{\mu_B^2}}|{\psi_p^E}\ran  = 0 $ except when R is an integer\footnote{This is because we can write the integral as $\lan{\psi_o}|{(\cp\zp)}^{{\f{n}{R} }-1}|{\psi_o}\ran$ which following the analysis of Appendix-A contributes only at pole.}. However for R an integer it contributes  a constant term independent of $\phi,E$, in the partition function and can be ignored by subtracting out an overall constant from the hamiltonian which amounts to multiplying the partition function by an overall factor. For the macroscopic loop operator in any other sector, we can proceed in the same way . So we can write the partition function as 
\ber
{\prod_E} \lbr 1 
&+& \lan{\psi_p^E}| (1+ {\frac{  \hcp\hzm+ \hcm\hzp }{\mu_B^2}})e^{{-2\beta}{\pi R}(\mu+E )})     |{\psi_p^E}\ran\rbr \n
&=& {\prod_E} \lbr1+\lan{\psi_p^E}| e^{{\rm log}(1-  2{\frac  {H_0} {\mu_B^2}})}e^{{-2\beta}{\pi R}(\mu+E )})     |{\psi_p^E}\ran\rbr\n
&=& {\prod_E}\lbr1+\lan{\psi_p^E}|  e^{{-2\beta}{\pi R}(\mu+ H -{\f{1}{2{\pi}{\beta}R}}{\rm log}(1-2{\frac{ H_o} {\mu_B^2}} )}    |{\psi_p^E}\ran \rbr\n
&=& {\prod_E} \lbr1+\lan{\psi_p^E}|  e^{{-2\beta}{\pi R}(\mu+{H_o^\prime}+{H_p})}    |{\psi_p^E}\ran \rbr\n
&=& {\rm Tr}_ {\psi_p^E} \lbr e^{{-2\beta}{\pi R}(\mu+{H_o^\prime}+{H_p})}\rbr  
 \la{reln}
\eer
where  $H_o^\prime$ is as discussed in (\ref{def2}), given by ${H_o^\prime} = {H_o} - {\f{1}{\tpr\beta}}{\rm log}(1-{\f{2H_o}{\mu_B^2}})$;\, ${H_p}={H_p}(\cpm\zpm, H)$ is the effective perturbation in the presence of momentum modes \footnote{ In order to reach from 2nd to 3rd step we used the same tricks of section 3 which implies that around a delta function in time, we can make the time interval infinitesimally small so that we can ignore the commutator terms ($\lbr {H_o^\prime},{H_o}\rbr$ +....higher commutators)  what can arise on exponential as a consequence of Baker Hausdorff formula }.  

     First note that $|{\psi_p^E}\ran $, the eigenstate of $H= {H_o}+{H_p}$ does not diagonalize the complete effective hamiltonian $H_{\rm eff}= H -{\f{1}{2{\pi}{\beta}R}}{\rm log}(1-2{\frac{ H_o} {\mu_B^2}})$. 

However in order to evaluate the partition function   we will follow (\ref{partition3}). This partition function is exactly given by the one with a shift  
$\psi^{\prime E}_p  = \CW ^\prime{\psi_o}(E)\,\, \Rightarrow\,\, \psi_p^{E^\prime} =\CW{\psi_o}(E^\prime) $ and the perturbing phase $\phi_{wp}(E) \rightarrow \phi(E(E^\prime))= \phi^{\prime}(E^\prime)$  evaluated w.r.t the effective hamiltonian $H={H_o}+ {H_p}(H,\cpm\zpm)$ but  without insertion of the macroscopic loop operator W,  where $\psi^{\prime E}_p$ is the basis which diagonalizes the deformed hamiltonian ${H_o^\prime}+{H_p}$ , as we discussed.  So following (\ref{partition3}) we can evaluate the partition function (\ref{reln})in the shifted basis 
\ber
{\psi_p^{\prime E}}
&\rightarrow & 
{\psi_p^{E^\prime}}={\cal{W}}_s {\psi_o^{E^\prime} }= e^{ {\f{1}{2}}\phi (E({E^\prime})) + R\sum_{k\geq 1} t_{\pm k} {(z_\pm\bar z_\pm)}^{k/R} + \sum_{k\geq 1} {1\over k} v_{\pm k}(E({E^\prime})) {(z_\pm \bar z_\pm)}^{-k/R} }\n
 {\psi_o^{E^\prime }}=
&            &  e^{ {\f{1}{2}}(\phi^\prime)({E^\prime}) + R\sum_{k\geq 1} t_{\pm k} (z_\pm\bar z_\pm)^{k/R} -
R \sum_{k\geq 1} {1\over k} {v^\prime}_{\pm k}({E^\prime}) (z_\pm \bar z_\pm)^{-k/R} } {\psi_o^{E^\prime}},
\la{result}
\eer
where the shifted dressing operator is given by
$${\cal{W}}_s = e^{ {\f{1}{2}}\phi (E({E^\prime})) + R\sum_{k\geq 1} t_{\pm k} (z_\pm\bar z_\pm)^{k/R} + \sum_{k\geq 1} {1\over k} v_{\pm k}(E({E^\prime})) {(z_\pm \bar z_\pm)}^{-k/R} }$$
and $\phi(E)$ is the phase for perturbed wave function (\ref{varphi}).
So following (\ref{expressiondual1}) we have the free energy given by
\ber
{\cal{F}}(\mu,R) &=& \phi(E({E^\prime} = {\f{ir}{\b R}}-\mu))\n
                 &=& -i {\sum _{r=n+ {\f{1}{2}} \ge 0}}{\phi^\prime}( {\f{ir}{\b R}}-\mu),
\label{expressiondualmod}
\eer
where we have
\be
{\phi^\prime} ({E^\prime})={\phi} (E)
\la{phirelationmod}
\ee

So we see the partition function in  the presence of D brane in a  background perturbed by momentum modes with compactified time is  obtained from the one without D brane by the  shift
$$ E \rightarrow {E^\prime} \quad;\quad  {\cal{W}} \rightarrow {\cal{W}}_s $$
which defines a deformed Fermi surface..
 \subsection{ Lax formalism for Type 0A MQM in the presence of D brane}
 Our lesson from the previous discussion is that Toda structure for  Type 0A  MQM perturbed by tachyonic modes ,in the presence of D brane can be obtained when we replace  
\ber
{\cal{W}}
& \rightarrow &  {\cal{W}}_s =e^{ iR\sum_{n\ge 1} t_{\pm n} \omega^{n/R}}\ 
e^{\mp  i{\phi^\prime}(E^\prime)}\ 
e^{ iR\sum_{n\ge 1}{ v_{\pm n}^\prime}(E^\prime   )\  \omega^{-n/R}}.\n
\psi_o^{\prime E} 
& \rightarrow & \psi_o^{E ^\prime} = {\psi_o}{(E- {\frac{1}{2\pi R}} \log(1-{\frac{2E}{\mu_B^2}} ))},
\la{newdress}
\eer
\ber
{L^\prime}_+ 
&=& {{\cal{W}}_s}_+ \omega
   {{\cal{W}}_s}^{-1}_+, \quad L_- = {{\cal{W}}_{s-}} \omega ^{-1}
  { {\cal{W}}_{s-}}^{-1}_-,\n
{M^\prime}_+ 
&=&  { {\cal{W}}_{s+}} {\hat{E}}
    { {\cal{W}}_{s+}}^{-1}, \quad M_- =  { {\cal{W}}_{s-}} {\hat{E}}
    { {\cal{W}}_{s-}}.
\la{lprm}
\eer
Note the operator algebra (\ref{laxcommutator}) remains same.
\be
\langle E |e^{\pm i{\phi^\prime}}{\hat {\cal{W}}_{s\pm}} 
 {L^\prime}_\pm |\cpm\zpm\rangle =\langle E |e^{\pm i{\phi^\prime}_0}\hat {\cal{W}}_{s\pm } \hcpm\hzpm
 |\cpm\zpm\rangle ,
\la{prm}
\ee
where 
${\phi^\prime}= \phi({E^\prime})$.  Expression of ${M^\prime}$ is as described in (\ref{mla})
\ber
\langle E |
& & e^{\pm i{\phi^\prime}} \hat \CW_\pm \; 
 {{M^\prime}_\pm }|\cpm\zpm\rangle
= \pm i (\zpm\p_{\zpm} + \cpm\p_{\cpm}+ 1)
\Psi^{E^\prime}_\pm (\cpm\zpm)\n
 &=& \left(\sum _{k\ge 1} 
 k \ t_{\pm k} \ {(\cpm\zpm)}^{ k/R}  +{E^\prime}+
\sum_{k\ge 1}{v^\prime}_{\pm k} \
 {(\cpm\zpm)} ^{-k/R}\right)\Psi^{E^\prime}_\pm ((\cpm\zpm) ).
\la{mexpression}
\eer
As the partition function described in terms of Fermi level ${\mu}$
\be
{M^\prime_\pm} =
\sum _{k\ge 1}  k t_{\pm k}   {L^\prime}_\pm^{ k/R}+\hat{\mu}+\sum_{k\ge 1} {v^\prime}_{\pm k}  {L^\prime_\pm}^{-k/R}.
\la{mpm}
\ee
 The structure of the integrable flow remain same.  The Toda flow equation will be given by
${\phi^\prime}(\mu)
\equiv \phi({E^\prime}=-\mu)$
\be
i{\p \over \p t_1}
{\p \over \p t_{-1}} {\phi^\prime}(\mu)
= e^{i{\phi^\prime}(\mu)-i{\phi^\prime}(\mu-i/R)}
 - e^{i{\phi^\prime}(\mu+i/R)-i\phi(\mu)}.
\la{prmphiflow}
\ee
Now in order to see that partition function is a tau function of Toda lattice hierarchy first note that
\be
{\cal{Z}}(\mu,t)=\prod\limits_{n \ge 0}
\exp\left[ {i\b}\phi\left(i{\f{1}{\b}}{n+\hf\over R}-\mu\right)\right].
\la{fren}
\ee
with Fermi level ${E^\prime}=-\mu$.
Now on the other hand, the zero mode of the perturbing phase
is actually equal to the zero mode
of the dressing operators (\ref{newdress}). Hence it is expressed through
the $\tau$-function as in (\ref{taur}).  Since the shift in the discrete
parameter n is equivalent to an imaginary shift of the chemical potential
$\mu$, so (\ref{fren}) implies
\be
e^{{i\b} \phi(-\mu)}= {\f { {{\cal{Z}}} (\mu + {\f{i}{2R\b}}   )   } {  {{\cal{Z}}} (\mu - {\f{i}{2R\b}}   )      }   }.
\la{phitau}
\ee
However from (\ref{taur}) we have
\be
e^{{i\b} \phi(-\mu)}= {\f { {\tau_o} (\mu + {\f{i}{2R\b}}   )   } {  {\tau_o} (\mu - {\f{i}{2R\b}}) }}
\la{phitau}
\ee
So  one concludes that
\be
{\cal{Z }}(\mu,t)=\tau_{0}(\mu,t).
\la{Ztauu}
\ee
 \subsection{Representation in terms of a  bosonic field}
Here we will study the classical limit following the analysis of \cite{intflow}
The momentum modes can be described as the
oscillator modes of a bosonic field $\varphi(\cp\zp,\cm\zm)
= \vp_+(\cp\zp)+ \vp_-(\cm\zm)$.
The bosonization formula is 
\be
\Psi^{{E^\prime}=-\mu-i} _\pm (\cpm\zpm)= {\cal{Z}}^{-1}
e^{\pm  i \vp_\pm (\cpm\zpm)}\cdot  {\cal{Z}}.
\la{boson}
\ee
(Note here in the presence of FZZT brane ${\mu}$ corresponds to the deformed Fermi surface)
where ${\cal{Z}}$ is the partition function and 
\be
\vp_\pm(\cpm\zpm) =  
+R \sum _{k\ge 1} t_k {(\cpm\zpm)}^{k/R} +
 {1\over R} \p_\m  +
\mu\log \cpm\zpm -R \sum_ {k\ge 1} 
{1\over k} {(\cpm\zpm)}^{-k/R}{\p\over \p t_k}.
\la{bosoni}
\ee
Then from (\ref{mexpression}) the operators $M_\pm$ are represented by
the currents $\cpm\zpm \p_\pm\vp$ 
\be
{M_\pm^\dag} \Psi _\pm ^E (\cpm\zpm)|_{E=-\mu-i} 
=  {\cal{Z}}^{-1}  \cpm\zpm \p_\pm \vp \cdot  {\cal{Z}}.
\la{rpr}
\ee
 \subsection{The dispersionless (quasiclassical) limit}
We consider the quasiclassical limit $\beta \rightarrow \infty$.  In this limit the integrable
structure described above reduces to the 
dispersionless Toda hierarchy
where the operators $\hat{\mu}$ and
$\hat{\omega}$ can be considered as a pair of 
classical canonical variables
with Poisson bracket
\be
\{  \o, \mu  \} = \o
\la{poison}
\ee
Similarly, all operators become $c$-functions 
of these variables.
The Lax operators can be identified
with the classical phase space 
coordinates $\cpm \zpm$, which 
satisfy
\be
\{ \cp, \zm\}=\{ \zp, \cm \}=1
\la{classcom}
\ee
The shape of the Fermi sea is 
determined by the  classical trajectory 
corresponding to the Fermi level ${E^\prime}=-\mu$.  So we have

\be
\cp\zm +\zp\cm  -{\f{1}{\tpr\b}}\log (1- {\f{2(\cp\zm +\zp\cm)}{\mu_B^2}}) - \log{\epsilon}= -\mu.
\la{fermishape}
\ee

Where $\log{\epsilon}$ is the cut-off cancelling the singular contribution from the  point 
$(1- {\f{2(\cp\zm +\zp\cm)}{\mu_B^2}}) =0$.  In the perturbed theory the
classical trajectories are of the form
\be
\cpm \zpm= {L^\prime}_\pm (\o, \mu).
\la{ldeformed}
\ee
where the functions $L_\pm$ are of the form 
\be
{L^\prime}_\pm (\o, \mu)= e^{{\frac{1}{2}} \p_\mu{\phi^\prime} }\ 
\o^{\pm 1}
\left(1+\sum_{k\ge 1} {a^\prime}_{\pm k}(\mu)\ 
\o^{\mp k/R}\right).
\la{lde}
\ee
The flows $H_n$ become  Hamiltonians
for the evolution along the `times'
$t_n$. The unitary operators $\CW_\pm$
becomes a pair of canonical transformations 
between the variables $\o, \mu$ and
$L_\pm, M_\pm$. Their generating functions are 
given by the 
 expectation values 
$S_\pm = {\cal{Z}}^{-1}\  \vp_\pm(\cpm\zpm)\ \cdot 
{\cal{Z}}$ of the
chiral components of the bosonic field $\phi$
\be
S_\pm = \pm R\sum _{k\ge 1}   t_{\pm  k} \ 
 {(\cpm\zpm)}^{ k/R}  +\mu \log (\cpm\zpm) -  {\phi ^\prime} \pm
R\sum_{k\ge 1} {1\over k} 
 {v_k^\prime}\   {(\cpm \zpm)}^{-k/R},
\la{s}
\ee
where $v_k = \p \CF/\p t_k$.
The differential of the function $S_\pm$ is
 \be
{dS}_\pm = M_\pm d{\rm log }(\cpm\zpm) + {\rm log} \o \ d\mu +
R \sum_{n\ne 0} {H_n} {dt_n}.
\la{sa}
\ee
If we consider 
the coordinate $\o$ as a function 
of either $\cp\zp$ or $\cm\zm$, then
\be
\o = e^{{\p_\mu} {S_+}(\cp\zp)} =e^{{\p_\mu}{S_-}(\cm\zm)}.,
\la{pm}
\ee
The classical string equation
\be
\cp\zm + \zp\cm  - {\f{1}{2 \pi  R}}\log(1-{\f{2(\cp\zm + \zp\cm)}{\mu_B^2}})= {M_+ }= {M_-},
\la{M}
\ee
can be written as
\ber
\cp\zm 
&+& \cm\zp- {\f{1}{2 \pi R}}{\rm log}(1-{\f{2(\cp\zm + \zp\cm)}{\mu_B^2}})\n
&=& \sum _{k\ge 1}  k t_ k 
 {(\cp\zp)}^{ k/R}  +\mu  +
\sum_{k\ge 1} v_ k{(\cp\zp)} ^{-k/R}.
\la{pp}
\eer
\section{Conclusion}
\setcounter{equation}{0}

Here we have studied Type 0A matrix model in the presence of spacelike D brane which are localized in matter direction. In  matrix model this is expressed by insertion of an operator $e^{W(t_o)}$ into the path integral. When we studied the respective MQM we found by application of Ward identity that the time translation invariance of the path integral in the presence of such operator gives the signal of nonconservation  of MQM hamiltonian. However we have shown in section 1.2 that this has a meaning that string hamiltonian is not being conserved in the presence of D brane which implies that string is going to off shell in the presence of D brane, which is never feasible.
In order to obtain right string theory picture we impose a constraint (\ref{impose}) on matrix model path integral in the presence of D brane. We explained that this condition has an effect to constrain the Hilbert space generated by macroscopic loop operator while keeping type 0A MQM unaffected. We have shown that when we impose the constraint we get the matter one point function from collective field theory.  We have further shown that exactly at the point of insertion of the brane ( which in string theory correspond to the point where open string ends are localized) the wave function for the right and left moving component of boundary state with any momentum appears to be identical which can be seen in matrix model as a consequence of this constraint. We also found right transition amplitude from a free fermionic state to coherent state.
   Next we consider type 0A MQM with the time t compactified on a circle. We have shown matrix model path integral in the presence of Dbrane can be expressed as Fredholm determinant.  We evaluated the thermal partition function in grand canonical ensemble. As the theory is defined on a circle so the partition function correspond to that of a deformed Fermi surface. We have further shown that in absence of any such constraint, the partition function diverges. 
     Finally we considered type 0A MQM in the background of momentum modes. First in section 4 we made a semiclassical analysis, studied fermionic scattering in the presence of D brane. We found the effective hamiltonian in the perturbed background from semiclassical analysis. We derived the grand canonical partition function in the perturbed background in the presence of D brane.  We have shown the partition function corresponds to tau function of Toda hierarchy. We have also analyzed the theory in dispersionless limit. 
\vskip 0.5mm
Its interesting to study T duality between type 0A and type 0B MQM in the presence of D brane.  One can also study the theory in the presence of flux background and see the consequence of the constraint.

\vskip15mm
\begin{center}
{\bf Acknowledgments}
\end{center}

The author wishes to express her deep sense of gratitude to Raghava Varma for
constant support, motivation and encouragement to pursue the research at IIT
Bombay. Its a pleasure to thank Satchidananda Naik for useful discussions
which lead to the finding of this problem and valuable suggestions during the
progress of the work. The author is  greatly indebted to P. Ramadevi for her generous
support and encouragement, necessary advices, and comments on the draft. The author
is  grateful to Sunil Mukhi and Koushik Ray for valuable discussions and
important comments. The author would also like to thank S. Uma Sankar and Soumitra
Sengupta for their inputs. 
Special thanks to Ankhi Roy and Rajeeb R. Mallick for their sincere cooperation
during the work in IIT Bombay. The author is also thankful to Partha Pratim Pal and
Colina Dutta for proof-reading the manuscript. Finally, the author wants to thank all
the research scholars in the High energy Physics group at IIT Bombay, in particular,
Reetanjali Moharana, Sushant k. Raut, Sasmita Mishra,Amal Sarkar,  Ravi Manohar, Neha Shah, Himani Bhatt, Kabita Chandwani, Suprabh Prakash for discussions and
 help. This work is supported by funds from IRCC, IITB and the research
development fund of Prof. Raghava Varma.

\vskip5mm
\appendix{\noindent\bf\Large{Appendix} {}}
\section{Appendix}
\setcounter{equation}{0}

Here we will show that type 0A MQM wave functions in the presence of D brane satisfy orthogonality and biorthogonality conditions.

\subsection{Orthogonality Condition}

The wave function is expected to show orthonormal property.
\\
1. For $t<{t_o}$ we have the wave function ${\psi_o}$ given in (\ref{freesolution}).  One can check the orthonormality property by considering the contour integral ,and it is given by \cite{yin}
\begin{equation}
\langle{\psi_{E^\prime}}|{\psi}_E\rangle ={\delta}(E-{E^\prime}).
\label{orthonormal}
\end{equation}
For ${t>{t_o}}$, first consider the wave function ${\psi_+}({z_+},{\overline{z}_+},t;E)$.  we have
\begin{eqnarray}
\langle{\psi}({E^\prime},t){\psi}(E,t)\rangle
&=& \int d{\zp }d{\cp} 
\lbrack\lbrace1-{{\rm log}(1+{\frac{  W (\cm\zm,2{H_o} )}{\mu_B^2}})\rbrace{e^{i{E^\prime}(t-{t_o})}}
{e^{i{ {\f{\phi_o}{2}} }(E^\prime)}}{( \cm\zm)}^{-i{E^\prime}-{\frac12}}}\rbrack\nonumber\\
& &\lbrack \lbrace 1- {\rm log}(1+ {\frac{  W(\cm\zm,2{H_o})    }{\mu_B^2}} )\rbrace{e^{-iE(t-{t_o})}} {e^{-i{\f{\phi_o}{2}} (E)}}
{(\cp\zp)}^{iE-{\frac12}}\rbrack.
\label{ort}
\end{eqnarray}
Now in order to see the orthonormal property first recall the commutation relation (\ref{com})

Note that ${\hat{z}_+}$ and ${\hat{z}_-}$  shifts E by -i and +i respectively.So we can write 
\begin{equation}
{\hat{\overline{z}_+}}{\hat{z}_+}={e^{\frac{-i\phi_o}{2}}}{e^{-i{\partial_E}}}{e^{\frac{i\phi_o}{2}}}
\label{express}
\end{equation}
Similarly 
\begin{equation}
{\hat{\overline{z}}_-}{\hat{z}_-}={e^{\frac{-i\phi_o}{2}}}{e^{i{\partial_{{E}}}}}{e^{\frac{i\phi_o}{2}}}.
\label{relation}
\end{equation}
Now to evaluate (\ref{ort}) first consider the expression 
\begin{eqnarray}
\langle{\psi_{+}}|\lbrack{\hat{\overline{z}}_-}{\hat{z}_-}{\rbrack^m}|{\psi_{+}}\rangle
&=& \int d{z_+ }d{\overline{z}_+} \lbrace{e^{i{E^\prime}t}}{e^{{\f{i}{2}}{\phi_o}(E^\prime)}}{{({z_+}{\overline{z}_+})}^{-i{E^\prime}-{\frac12}}}\rbrace\n
& & \lbrack\pcp\pzp{\rbrack^m}\lbrace{e^{-iEt+mt}} {e^{-{\f{i}{2}}{\phi_o}(E-mi)}}
{{({z_+}{\overline{z}_+})}^{iE-{\frac12}}}\rbrace.
\label{ortho}
\end{eqnarray}
For $E\ne{E^\prime}$ this can just be written as
\begin{eqnarray}
\langle{\psi_{+}}|\lbrack{\hat{\overline{z}}_-}{\hat{z}_-}{\rbrack^m}|{\psi_{+}}\rangle
&=& {e^{-{\f{i}{2}}{\phi_o}}}{e^{im{\partial_{E}}}}
{e^{{{\f{i}{2}}\phi_o}}}\int d{z_+ }d{\overline{z}_+} \lbrace{e^{i{E^\prime}t}}{e^{{\f{i}{2}}{\phi_o}(E^\prime)}}{{({z_+}{\overline{z}_+})}^{-i{E^\prime}-{\frac12}}}\rbrace\n
& & \lbrace{e^{-iEt}} {e^{-{\f{i}{2}}{\phi_o}(E)}}
{{({z_+}{\overline{z}_+})}^{iE-{\frac12}}}\rbrace\nonumber\\
&=& {e^{-{\f{i}{2}}{\phi_o}}}{e^{im{\partial_{E}}}}
{e^{{\f{i}{2}}{\phi_o}}}\langle{\psi_{+}}({E^\prime})|{\psi_{+}}(E){\rangle_{E\ne{E^\prime}}}\nonumber\\
&=&  0 .
\label{orthogon}
\end{eqnarray}

For $E={E^\prime}$ (\ref{ort}) takes the form
\begin{eqnarray}
\langle{\psi_{+}}|\lbrack{\hat{\overline{z}}_-}{\hat{z}_-}{\rbrack^m}|{\psi_{+}}\rangle
&=& \lbrace{e^{i{\phi_o}(E)}}{e^{-i{\phi_o}(E-mi)}}\rbrace\int d{z_+ }d{\overline{z}_+} 
\lbrack{\overline{z}_+}{z_+}{\rbrack^{-m-1}}{e^{-mt}} .
\label{orthomod}
\end{eqnarray}
From contour integral which is 0 for $m\ge1$. Also we conclude that $\langle{\psi_{+}}|\lbrack{\hat{\overline{z}}_+}{\hat{z}_+}{\rbrack^m}|{\psi_{+}}\rangle$
and $\langle{\psi_{-}}|\lbrack{\hat{\overline{z}}_-}{\hat{z}_-}{\rbrack^m}|{\psi_{-}}\rangle $ diverge.  Here before going to show the orthogonality lets consider the situation when m is not an integer. This kind of integration we had in the section 5 in the expression $\lan{\psi_o}|{(\cp\zp)}^{{\f{n}{R}}-1} |{\psi_o}\ran$. We can consider this as the product over two branh cut integrals $\zp$ and $\cp$ and the each branch cut integral can be expressed as the sum of two standard contour integral with the cut on right and left side of the real axis $0 \geq x \geq \infty$ and $-\infty \geq x \geq 0$ respectively and with a pole at $x=0$. One can see that the integral turns out to be zero for any noninteger ${\f{m}{R}}$. We have nonzero contribution only when R is an integer and the contributing term is m=R.This term corresponds to a pure pole and give a constant contribution to the integral,
Now back to the question of orthogonality.
So using (\ref{relation} ,\ref {orthomod}), we can be written (\ref{ort}) as
\begin{eqnarray}
\langle{\psi}({E^\prime},t)|{\psi}(E,t)\rangle
&=& \int d{z_{+} }d{\overline{z}_+} 
\lbrack\lbrace1 - {\rm log}(1+{\frac{  f (0,2{H_o} )}{\mu_B^2}})\rbrace{e^{i{E^\prime}(t-{t_o})}}
{e^{{\f{i}{2}}{\phi_o}(E^\prime)}}{{({z_+}{\overline{z}_+})}^{-i{E^\prime}-{\frac12}}}\rbrack\nonumber\\
& &\lbrack \lbrace 1- {\rm log}(1+ {\frac{  f(0,2{H_o})    }{\mu_B^2}} )\rbrace{e^{-iE(t-{t_o})}} {e^{-{\f{i}{2}}{\phi_o}(E)}}
{{({z_+}{\overline{z}_+})}^{iE-{\frac12}}}\rbrack\nonumber\\
&=& \int d{z_{+} }d{\overline{z}_+}
\lbrack\lbrace1-{\rm log}(1+{\frac{  f (0,2E)}{\mu_B^2}})\rbrace{e^{i{E^\prime}(t-{t_o})}}
{e^{{\f{i}{2}}{\phi_o}(E^\prime)}}{{({z_+}{\overline{z}_+})}^{-i{E^\prime}-{\frac12}}}\rbrack\nonumber\\
& &\lbrack \lbrace 1-{\rm log}(1+ {\frac{  f(0,2E )  }{\mu_B^2}} )\rbrace{e^{-iE(t-{t_o})}} {e^{-{\f{i}{2}}{\phi_o}(E)}}
{{({z_+}{\overline{z}_+})}^{iE-{\frac12}}}\rbrack.
\label{orthogonal2}
\end{eqnarray}
  Now write ${\phi_o}= {\phi_{oRe}} + i{\phi_{oIm}}$.  Shifting ${\phi_{oIm}} \rightarrow {\phi_{oIm}}-{\f{i}{2}}{\rm log}\lbrack1-{\rm log}(1+ {\frac{  f(0,2E )  }{\mu_B^2}} )\rbrack$ we get the orthogonal property (\ref{orthonormal}) with
\begin{equation}
{\psi_+}(E,t)= {e^{iE(t-{t_o})}} {e^{-{\f{i}{2}}{\phi_{0+}}(E)}}
{{({z_+}{\overline{z}_+})}^{iE-{\frac12}}},
\end{equation}
where 
\begin{equation}
{\phi_{o+}}(E)={\phi_o}(E)-i {\rm log}\lbrack1-log(1+ {\frac{  f(0,2E   }{\mu_B^2}} )\rbrack.
\label{phi}
\end{equation}
Similarly for the wave function in ${z_-}$ representation we find 
\begin{equation}
{\phi_{o-}}(E)={\phi_o}(E)+ilog\lbrack1- log(1+ {\frac{  f(0,2E   }{\mu_B^2}} )\rbrack
\label{phi}
\end{equation}
So we see the consequence of the insertion of macroscopic loop operator is that the phase of the wave function develops an imaginary part which is associated with tunneling.

\subsection{Biorthogonality Relation}
The wave function is expected to satisfy the following biorthogonality condition which has a consequence in the evaluation of scattering amplitude 
\cite{intflow,yin,timedependentbgd}.
\begin{equation}
\int d{\zp}d{\zm}d{\cp}d{\cm}{\overline{\psi^E_+}}(\cp\zp,t){e^{i(\cp\zm + \cm\zp)}} {\psi^{E^\prime}_-}(\cm\zm,t) = {\delta}(E-{E^\prime})
\label{biortho}
\end{equation}
Now for $t\le{t_o}$ we have ${\psi}={\psi_o}$ and for which biorthogonality relation is already derived in \cite{intflow},\cite{yin}  giving $e^{i\phi_0(E)}
  =    {\f{\Gamma( iE + 1/2)}{\Gamma( iE + 1/2)}}$ \cite{yin}.  For $t\ge {t_o}$ , .Biorthogonality relation takes the form
\ber
\int d{\zp}d{\zm}d{\cp}d{\cm}
& &{\overline{\psi}^E_+}(\cp\zp,t)\lbrace1-{\hat{W}}({\hzp,\hcp,{H_o},t})\rbrace{e^{i(\cp\zm + \zp\cm)}} )
(1-{\hat{W}}({\hzp,\hcp,{H_o},t})){\psi^{E^\prime}_-} \n
&=& {\delta}(E-{E^\prime}),
\label{biortho2}
\eer
In order to show this note that 
\begin{eqnarray}
\int d{\zp}d{\zm}d{\cp}d{\cm}
& & {\overline{\psi^E_+}}({\zp},{\cp},t)\lbrace1-{\hat{W}}({\hzm,\hcm,{H_o},t})\rbrace{e^{i(\zp\zm + \cp\cm)}} )(1-{\hat{W}}({\hzm,\hcm,{H_o},t})){\psi^{E^\prime}_-} \nonumber \\
&=& \int d{\zp}d{\cp}{\overline{\psi^E_+}}({\zp},{\cp},t)\lbrace1-{\hat{W}}({\hzm,\hcm,{H_o},t})\rbrace (1-{\hat{W}}({{\f{\p}{\p\zp}},{\f{\p}{\p\cp}},{H_o},t}))\nonumber\\
& &  \int d{\zm}d{\cm}{e^{i(\cp\zm + \cm\zp)}} ){\psi^{E^\prime}_-}({\cm\zm},t) \nonumber \\
&=& \int d{\zp}d{\cp}{\overline{\psi^E_+}}({\zp},{\cp},t)\lbrace1-{\hat{W}}({\hzm,\hcm,{H_o},t})\rbrace (1-{\hat{W}}({\hzm,\hcm,{H_o},t}))\nonumber\\
& &  \int d{\zm}d{\cm}{e^{i(\zp\zm + \cp\cm)}} ){\psi^{E^\prime}_-}({\zm},{\cm},t) \nonumber \\
&=& R(E)\int d{\zp}d{\cp}{\overline{\psi^E_+}}({\zm},{\cm},t)\lbrace1-{\hat{W}}({\zm,\cm{H_o},t})\rbrace (1-{\hat{W}}({\zm,\cm,{H_o},t}))\nonumber\\
 {\psi^{E^\prime}_-} ({\zp},{\cp},t)\nonumber\\
 &=& {R}(E) {e^{i{\phi_{o+}} }}{\delta}(E-{E^\prime}).
 \label{biorthogonal}
 \end{eqnarray}
 Where in order to come from 2nd to 3rd step we used the fact that in ${z_+}$ representation we have $\hzm,\hcm = {\f{\p}{\p\zp}},{\f{\p}{\p\cp}}$
The integral in the 4th step, we have evaluated in (\ref{orthogonal2}) leads to the last step.  In order to get the biorthogonality relation(\ref{biortho2}) we must need to set $e^{i{\phi_{o+}}(E)}= R(E) $. Compared to the $t\le {t_o}$  case note the shift of ${\phi_o}(E)$ due to the insertion of macroscopic loop operator.
\begin{equation}
R(E){\psi^{E^\prime}_+}(\zp,\cp,t)={ \int _{-\infty}^{\infty}}d{\zm}d{\cm}{e^{i(\zp\zm + \cp\cm)}} ){\psi^{E^\prime}_-}(\zm,\cm,t)  
\end{equation}
R(E) getting absorbed to decide ${\phi_o}$ and shifting ${\phi_o}$ according to (\ref{phi}) we get the above result.

\end{document}